\documentclass[twocolumn]{aastex631}

\shorttitle{Solar System Hunting}
\shortauthors{Yahalomi et al.}
\graphicspath{{./}{figures/}}
\begin{document}

\title{Detecting Solar System Analogs through Joint Radial Velocity/Astrometric Surveys}

\correspondingauthor{Daniel A. Yahalomi}
\email{daniel.yahalomi@columbia.edu}

\author[0000-0003-4755-584X]{Daniel A. Yahalomi} 
\altaffiliation{LSSTC DSFP Fellow}
\affiliation{Department of Astronomy, Columbia University, 550 W 120th St., New York NY 10027, USA}

\author[0000-0003-4540-5661]{Ruth Angus}
\affiliation{American Museum of Natural History, Central Park West, Manhattan, NY, USA}
\affiliation{Center for Computational Astrophysics, Flatiron Institute, 162 5th Avenue, New York, NY 10010, USA}
\affiliation{Department of Astronomy, Columbia University, 550 W 120th St., New York NY 10027, USA}

\author[0000-0002-5151-0006]{David N. Spergel}
\affiliation{Center for Computational Astrophysics, Flatiron Institute, 162 5th Avenue, New York, NY 10010, USA}

\author[0000-0002-9328-5652]{Daniel Foreman-Mackey}
\affiliation{Center for Computational Astrophysics, Flatiron Institute, 162 5th Avenue, New York, NY 10010, USA}



\begin{abstract}

Earth-mass exoplanets on year-long orbits and cool gas giants (CGG) on decade-long orbits lie at the edge of current detection limits. The Terra Hunting Experiment (THE) will take nightly radial velocity (RV) observations on HARPS3 of at least 40 bright nearby G and K dwarfs for 10 years, with a target 1$\sigma$ measurement error of $\sim$0.3 m/s, in search of exoplanets that are Earth-like in mass and temperature. \textnormal{However, RV observations can only provide minimum mass estimates, due to the mass-inclination degeneracy. Astrometric observations of these same stars, with sufficient precision, could break this degeneracy}. Gaia will soon release $\sim$100-200 astrometric observations of the THE stars with a 10 year baseline and \textnormal{$\sim$34.2}\,$\mu$as 1$\sigma$ \textnormal{along-scan} measurement error.  The Nancy Grace Roman Space Telescope will be capable of precision astrometry using its wide field imager (target $\sim$5-20\,$\mu$as 1$\sigma$ measurement error for bright stars) and could extend the astrometric observational baseline to $\sim$25 years. We simulate and model an observing program that combines data from these three telescopes. We find that (1) THE RVs and Gaia astrometry can detect Earth-like and CGG-like exoplanets around bright Sun-like stars at 10 parsecs and that (2) adding Roman astrometry improves the detection precision for CGG masses and periods by a factor up to $\sim$10 and $\sim$4, respectively. Such a survey could provide insight into the prevalence of \textnormal{Solar System analogs, exoplanet architectures reminiscent of the mass and orbital separation hierarchy of our Solar System, for the nearest Sun-like stars.}


\end{abstract}



\section{Introduction} \label{sec:intro}

\subsection{Background}
The \textnormal{search for} and study of extrasolar planets (exoplanets) has a long history of exciting and surprising discoveries \citep{Latham1989, Wolszczan1992, Mayor1995, Correia2005, Borucki2010, Ricker2015, Butler2017}. Astronomers have detected and confirmed thousands of exoplanets \citep{NEA12}. Many of these worlds bear little similarities to our own Solar System.

The formation and migration of gas giant planets and the interactions between gas giants in our Solar System played a critical role in the characteristics of the planets in our Solar System \citep{Tsiganis2005, Morbidelli2005, Gomes2005, Walsh2011, Batygin2015}. Further, Jupiter and Saturn may have played a crucial role in Earth's habitability, in that they (1) help stabilize Earth's orbit and thus climate and (2) shield Earth from possibly devastating impacts \citep{Raymond2004}. Investigations of and searches for true Earth-twins should thus focus on \textnormal{systems we will refer to as} Solar System analogs: exoplanetary systems with an Earth analog (rocky 1 M$_\Earth$ exoplanet in the habitable zone) and a cool gas giant (CGG) on a Jupiter-like or a Saturn-like orbit around Solar type stars. As the Earth is the only presently known habitable planet, it is natural to search for Solar System analogs in pursuit of habitable exoplanets. Further, studies of the demographics of stars with both an Earth-like exoplanet and a CGG-like exoplanet will be of particular interest in the coming decades.

\begin{figure}
\centering
\includegraphics[width=0.45\textwidth]{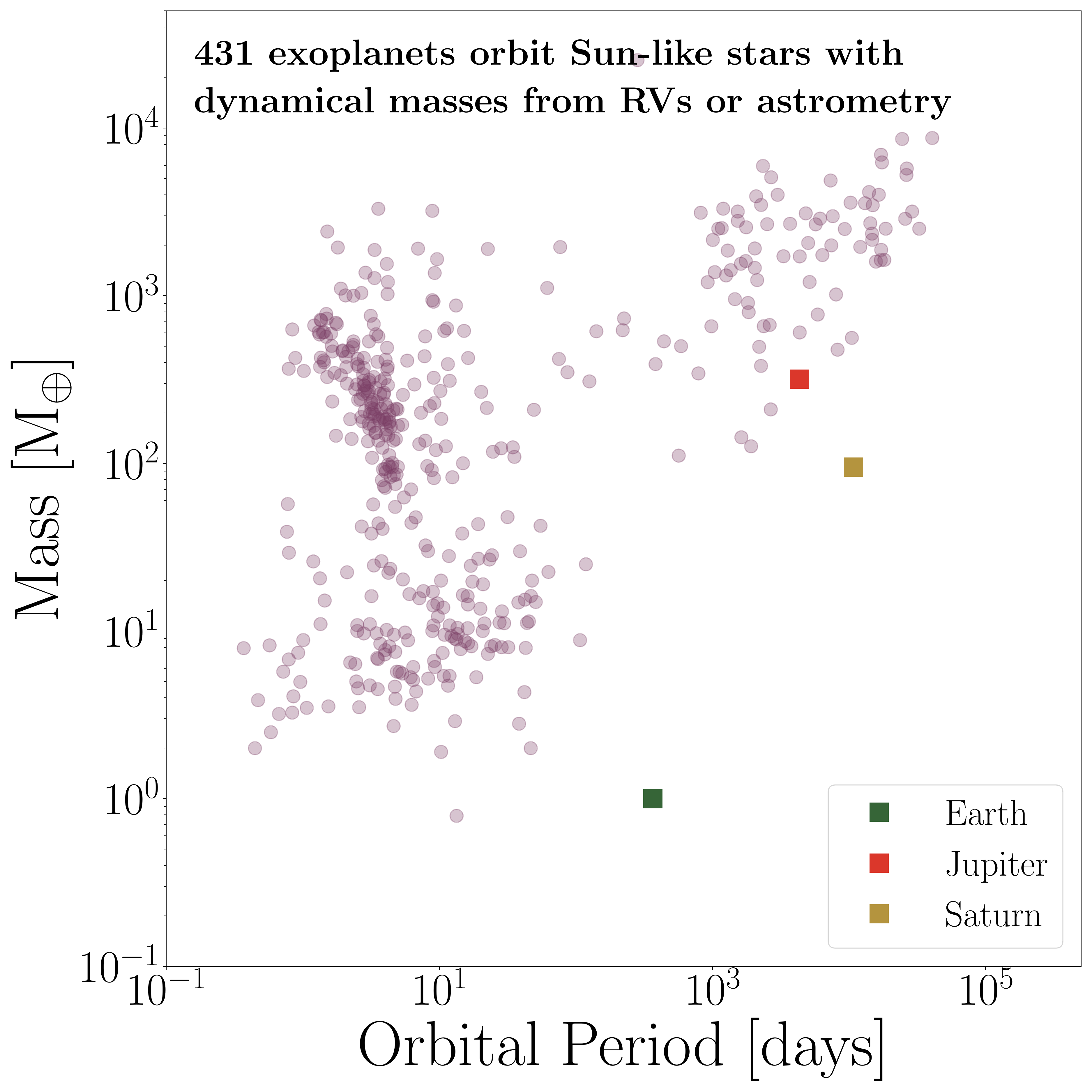}
 \caption{Mass vs. orbital period of confirmed exoplanets 
 with dynamical masses from radial velocity or astrometric detections orbiting Sun-like stars. We define a Sun-like star as any star with a mass and radius in the ranges [0.75-1.25 M$_\Sun$] and [0.75-1.25 R$_\Sun$], respectively. Earth, Jupiter, and Saturn are also plotted to show their relative location with confirmed exoplanets. The paucity of exoplanets at low masses and long periods reflects the limitations of our observational techniques rather than true exoplanet demographics. \textnormal{Exoplanet data extracted from the composite planetary systems dataset, using all systems without a controversial flag, from the \citet{NEA12}, accessed on 18 September 2023.}}

 \label{fig: exoplanet_demographics}
\end{figure}

Solar System analogs lie outside current telescope capabilities. Figure \ref{fig: exoplanet_demographics} shows the mass vs. period of all confirmed exoplanets around Sun-like stars, \textnormal{with dynamical masses from RV or astrometry, from the exoplanet archive\footnote{\url{https://exoplanetarchive.ipac.caltech.edu/}}} where we define a Sun-like star as any star with a mass and radius in the ranges [0.75-1.25 M$_\Sun$] and [0.75-1.25 R$_\Sun$], respectively. We can see that there have been no true Earth analog detections and that the CGG detections are quite rare. A driving factor in the demographics of known exoplanets is that the transit and radial velocity methods, workhorses in modern exoplanet detection, are both more sensitive to larger exoplanets on closer in orbits \citep{Borucki2010, Ricker2015, Butler2017}.

 \begin{figure*}[htb] 
    \centering 
    \includegraphics[width=\textwidth]{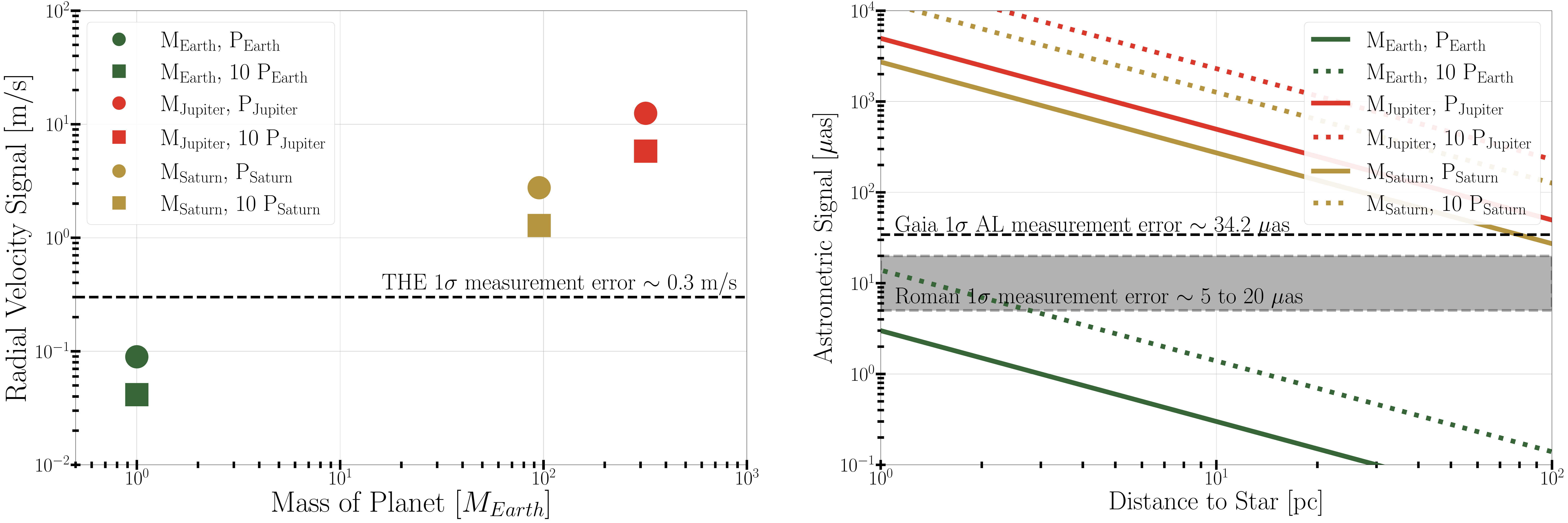}
    \caption{[$\textbf{Left}$] RV signal for Earth-mass, Jupiter-mass, and Saturn-mass objects on Earth-like, Jupiter-like, and Saturn-like eccentric orbits, respectively, around a Solar mass star with varying orbital periods. THE is pushing towards the 0.1 m/s detection limit for Earth-like exoplanets, with 10 years of nightly observations with target 1$\sigma$ measurement error of $\sim$0.3 m/s per observation. [$\textbf{Right}$] Astrometric signal for Earth-mass, Jupiter-mass, and Saturn-mass objects around Solar mass stars with varying periods on circular orbits. Gaia astrometry will have a $\sim$34.2 $\mu$as 1$\sigma$ measurement error per \textnormal{field-of-view passing for the along-scan (AL) direction} for bright stars. Roman astrometry has a target $\sim$5-20 $\mu$as 1$\sigma$ measurement error per observation for bright stars. With Gaia and Roman, the astrometric observations can have a baseline of up to $\sim$25 years. As the radial velocity signal increases with shorter orbital periods and the astrometric signal increases with larger orbital periods, we can see how the two observing methods are complementary in period space. We emphasise that these (target) measurement errors are not a detection threshold, but rather a per observation estimated measurement error.}
    \label{fig: signal_strength}
\end{figure*}

In the coming decade, with radial velocity surveys designed for the purpose of Earth-twin detection \citep{Hall2018} and with astrometric observations from space missions like Gaia, which will release epoch astrometry in Data Release 4 (DR4) \citep{Gaia2016} and The Nancy Grace Roman Space Telescope, set to launch in late 2027 \citep{Spergel2015}, there is promise in the detection of Solar System analogs. In addition to its well studied ability to contribute to exoplanet science via the microlensing method \citep{Penny2019, Johnson2020} and via the transit method \citep{Montet2017}, Roman will also be capable of high precision astrometric exoplanet detection \citep{Sanderson2019}. Astrometric observations complement radial velocity surveys targeting Earth-analogs in that (1) astrometric measurements can break the mass-inclination degeneracy inherent in radial velocity observations without the geometric and observational requirements inherent in transit observations, (2) astrometric measurements, which are more sensitive to longer periods, complement RV observations, which are more sensitive to shorter periods, and (3) astrometric observations could allow us to marginalize over stellar noise in the RV data as astrometry is less sensitive to stellar noise than the Doppler method \citep{Makarov2009, Shao2018}. The complementary nature between RVs and astrometry can be seen in Figure \ref{fig: signal_strength}. Here, we have assumed circular and face-on orbits. We plot RV amplitude vs. planet mass and astrometric amplitude vs. distance to the star for Earth-mass, Jupiter-mass, and Saturn-like objects with periods of 1x and 10x their Solar System values around the Sun. 

One thing to note with astrometric observations is that the detection capabilities of astrometric surveys drop very steeply when the period of the orbit extends past the duration of the astrometric observations \citep{Casertano2008}. Thus, in targeting exoplanets on several decade long orbits with astrometry, combining observations from Gaia and Roman, which could extend the astrometric baseline to $\sim$25 years, will be critical in extending the period baseline for detection capabilities.

In this paper we demonstrate how Gaia astrometry in combination with a targeted Roman astrometry program of \textnormal{Terra Hunting Experiment} (THE) stars could be used to (1) break the mass-inclination degeneracy present in radial velocity observations and (2) accurately detect both Earth-like and CGG-like exoplanets around nearby Sun-like stars. Further, we show that these Roman observations specifically will help in more precise determinations of both the periods and masses of cool gas giants. By combining observations from the THE Doppler survey with Gaia and Roman astrometry, we can investigate the probability that a Solar-type star has an orbiting CGG, given that the Solar-type star has an orbiting Earth-like exoplanet ( P(CGG $|$ Earth) ). Specifically, in this proposed Roman observing program, we will obtain follow-up astrometric observations of stars that show radial velocity signals consistent with an Earth analog via the Terra Hunting Experiment. In all \textnormal{of} our joint RV and astrometry models, we include Gaia astrometry, as it is reasonable to assume that Gaia will observe all THE stars, as the THE survey is selecting targets based on the criteria of being bright and nearby \citep{Hall2018}.

In combination with exoplanet demographic models, one could then infer, using Bayes theorem, the probability that a Solar-type star has an orbiting Earth-like exoplanet, given that the star has an orbiting CGG (P(Earth $|$ CGG)), by using the Equation \ref{eq: Bayes}:

\begin{equation} \label{eq: Bayes}
    P(\textrm{Earth} | \textrm{CGG}) = \frac{P(\textrm{CGG} | \textrm{Earth}) \, P(\textrm{Earth})} {P(\textrm{CGG})}
\end{equation}

This version of the equation has the inherent benefit, in that it is easier to detect cool gas giants, that are much more massive than Earth-like exoplanets, and thus have a larger gravitational effect on their host star. Admittedly, this version of the equation has the downside in that the model would be built on assumptions of P(Earth) and P(CGG). To date, similar studies have been conducted for super-Earths and cold-Jupiters \citep{Zu2018, Masuda2020} as well as for hot-Jupiters and cold-Jupiters \citep{Knutson2014}. 

\citet{Zu2018} investigated the relations between inner ($<$ 1 AU) super-Earths and outer ($>$ 1 AU) cold-Jupiters and determined P(\textrm{cold-Jupiter} $|$ \textrm{super-Earth}) $\sim$ 30 $\%$. By comparing this value with results from \citet{Cumming2008}, which predicts that 10$\%$ of Sun-like stars should have at least one cold-Jupiter, they determined that cold-Jupiters appear three times more often around super-Earth hosts than around Sun-like stars. Finally, by arguing that P(\textrm{super-Earth}) = 30$\%$ and using a version of Bayes theorem as shown in Equation \ref{eq: Bayes}, \citet{Zu2018} argued that P(\textrm{super-Earth} $|$ \textrm{cold-Jupiter}) $\sim$ 90$\%$. Subsequently, \citet{Masuda2020} showed that stellar systems with super-Earths and cold-Jupiters tend to be coplanar, but not quite as coplanar as the Solar System.

As precision astrometric observations have, to date, not been used to search for Solar System analogous exoplanets, we rely on simulated results to estimate the noise floor present from astrometric jitter. In a recent series of papers, \citet{Shapiro2021} and \citet{Sowmya2022} present a model of astrometric jitter of Sun-like stars. \citet{Shapiro2021} focused on observations from the ecliptic, while \citet{Sowmya2022} studied the dependence of astrometric jitter on inclination, metallicity, and active-region nesting for Sun-like stars in Gaia and Small-JASMINE \citep{Yano2013, Utsunomiya2014} data. They found that the photocenters of stars observed out of their equatorial planes experience systematic shifts over the activity cycle. This shift in photocenter could introduce some astrometric jitter to the data. In general, \citet{Sowmya2022} showed that photocenter displacements in the Small-JASMINE infrared filter are smaller than those in the Gaia-G filter. As Roman will observe down to K-band \citep{Spergel2015}, it would likely have less astrometric jitter than Gaia as well. \textnormal{Further, it should be noted that it was recently showed that the precision of Gaia astrometry is insufficient to detect starspot induced jitter from stars with near-solar activity levels \citep{Morris2018}. Therefore, we should expect that the the level of astrometric jitter from activity cycles of Sun-like stars should not affect the Gaia data in general.}

\subsection{Nancy Grace Roman Space Telescope}

The Nancy Grace Roman Space Telescope is set to launch in 2027.\footnote{\url{https://roman.gsfc.nasa.gov}} Roman will be used to investigate a wide range of astronomy topics, including exoplanet detection and imaging. Roman will have a 2.4m primary lens, which is the same size as the Hubble Space Telescope, but a viewing window 100 times larger than Hubble. Roman will have no proprietary time and all observing time will be competed via external proposals. Roman has a 5-year primary mission length and will hopefully operate for at least 10 years \citep{Spergel2015}.

As previously mentioned, Roman is expected to detect thousands of exoplanet microlensing events, which will be crucial in expanding the baseline of period sensitivity in exoplanet detection and improving our understanding of long period exoplanets \citep{Penny2019, Johnson2020}. However, the microlensing method only provides a single observation of an exoplanetary system. Roman will also add to the growing list of exoplanet transit detections, with current estimates predicting order 100,000 new exoplanet detections \citep{Montet2017}. 

Roman will also be capable of high precision astrometry and will provide at least a factor of three improvement over the current state of the art astrometric observations in this wavelength range. It has been shown that with current specs, Roman astrometry of bright nearby stars ($<$ 10 parsecs) could detect Earth-mass exoplanets in some cases in the habitable zone \citep{Melchior2018, Sanderson2019}, as well as more distant exoplanets with periods up to $>$10 years by adding earlier astrometric observations from Gaia \citep{Perryman2014, Gaia2016}.
It is estimated that Roman will be capable of single observation 1$\sigma$ astrometric precision of $\sim$10 $\mu$as \citep{Sanderson2019}. 

As the stars we would like to observe are very bright, exoplanet astrometry with Roman can be pursued using two different methods, in order to avoid saturation:
\begin{enumerate}

    \item \textbf{Diffraction Spike Method:} In the diffraction spike method, observations are centered on the detector's diffraction spike. This is possible as Roman uses H4RG detectors rather than CCDs, in which pixels bleed into nearby pixels when saturated. Measurement accuracy with this method will likely be limited by systematics, and so it is ideal to take multiple observations of a system to reduce uncertainties \citep{Sanderson2019}. \citet{Melchior2018} investigated the capabilities of a Roman dedicated guest observing plan using the diffraction spike method for detection of Earth-mass exoplanets. In the paper, they argued that Roman astrometry using the diffraction spike method could obtain 10 $\mu$as astrometric observations with a single 100s exposure for a R=6 or a J=5 star -- precise enough to detect Earth-mass exoplanets on $\geq$ 1 year orbits around stars within a few parsecs.
    
    \item \textbf{Spatial Scanning Method:} In the spatial scanning method, as the name suggests, the telescope is slowly slewed in a specific direction during the observation. This creates spatial tracks for the brighter stars and allows for the observation of reference stars in the same field. In so doing, this observational technique avoids saturation by spreading the signal over hundreds, or even thousands, of pixels -- and thus allows for orders of magnitude of more photons. Subsequent spatial scans in two perpendicular directions allows for three-dimensional high precision observations. Similarly to the diffraction spike method, measurement accuracy with spatial scanning will likely be limited by systematics, and so it is ideal to take multiple observations of a system to reduce uncertainties \citep{Sanderson2019}. The spatial scanning method has been successfully used with Hubble observations to study bright Cepheids with astrometric precision of 20-40 $\mu$as \citep{Riess2014}. The 100x larger Roman field-of-view should enable much higher precision observations than the Hubble astrometric observations via spatial scanning.

\end{enumerate}

\subsection{Gaia}
Gaia is a space-based telescope with the mission to chart a three-dimensional map of the Milky Way, in pursuit of understanding the formation, composition, and evolution of the Milky Way.\footnote{\url{https://www.cosmos.esa.int/web/gaia}} Gaia, which launched in 2013, had a nominal mission length of 5 years, but it has been extended \textnormal{until the exhaustion of its cold gas propellant in the second quarter of 2025.} Gaia will release individual epoch astrometric observations in Gaia Data Release 4 (DR4), and is expected to have a significant impact on the knowledge of both exoplanet demographics and physical properties. The utility of Gaia for exoplanet detection and characterization via the astrometric method has been well studied, and current models and simulations predict the detection of \textnormal{as many as tens of thousands of} exoplanets for a 5 year mission \citep{Bernstein1995, Casertano1996, Lattanzi2000, Casertano2008, Perryman2014, 
Sozzetti2014, Ranalli2018}. Gaia is estimated to have 1$\sigma$ astrometric precision of \textnormal{$\sim$34.2 $\mu$as for the brightest nearby stars (for the  along-scan accuracy per field-of-view passage)} \citep{Perryman2014}. Gaia will make between 100 and 200 field-of-view passages for most stars in its observing field \citep{Perryman2014}.  By combining astrometric data from Gaia and Roman we can extend the baseline of observations, thus extending sensitivity well past 10 years.

\textnormal{Gaia, as for its predecessor Hipparcos, continuously scans the sky in order to provide maximal sky coverage. Specifically, Gaia's ``scanning law'' was created to maintain a constant (solar) thermal payload illumination, while maximizing separability of the astrometric parameters \citep{Perryman2014}. We can thus expect that data from future Gaia Data Releases will consist of  along-scan abscissa times series measurements with respect to a reference position ($\alpha_0$, $\delta_0$) together with the associated scan angles and parallax factors, similar to Gaia Data Release 3 (DR3) \citep{Holl2023}. Due to this scanning law, astrometric observations from Gaia must be separated in the along-scan (AL) and across-scan (AC) directions. It is important to note that Gaia precision will be $\sim$5 times worse in the AC direction than in the AL direction \citep{Lindegren2012, Abbas2017}.}

Data from Gaia Data Release 3 (DR3) \citep{Gaia2022a}, has led to several recent developments and discoveries in exoplanet astrometry. Assuming a Solar-mass host, \citet{Holl2023} presented a catalog of 17 (9 validated) planetary-mass,  52 (29 validated) brown dwarf regime, as well as 1093 (160 validated) low mass stellar companion regime candidates from a exoplanet search pipeline run on Gaia DR3 astrometric data. Here the label validated means that the values are consistent with other data in the literature. The \citet{Gaia2022b} presented 11 specific astrometrically detected exoplanets with high-precision radial velocity data in the literature. For these 11 exoplanets, they compared the astrometric orbital solutions with the published radial velocity orbital solutions and demonstrated that the periods and eccentricities of these two independent solutions were consistent. \citet{Winn2022} subsequently presented joint radial velocity and Gaia DR3 astrometric constraints on exoplanetary orbits for three exoplanets and presented inconsistencies for 4 other exoplanetary systems (1 of which is likely caused by additional exoplanets not included in the astrometric model, 2 of which are likely caused by inaccurate treatment of non-Gaussian uncertainties in the Gaia orbital solutions, and 1 of which is currently unexplained). As Gaia DR3 did not include time-series epoch astrometry data, \citet{Winn2022} generated posterior probability distributions from the Gaia DR3 two-body tabular results, for each of the RV orbital parameters. These probability distributions were then used in the joint astrometric and RV fits. 

This early exoplanet astrometry work from Gaia DR3 should be only the tip of the \textnormal{iceberg of astrometric detections of exoplanets}, once time-series epoch astrometry is released in Gaia DR4.

\subsection{Terra Hunting Experiment}
The Terra Hunting Experiment (THE) will be conducted on the HARPS3 instrument, a fibre fed, high resolution, high stability, echelle spectrograph.\footnote{\url{https://www.terrahunting.org}} HARPS3 will be installed on the 2.5m Isaac Newton Telescope in La Palma in the Canary Islands \citep{Thompson2016, Hall2016}. HARPS3 will be a close-copy to HARPS \citep{Mayor2003, Rupprecht2004} and HARPS-North \citep{Cosentino2012}. THE aims to discover Earth-twins via 10-years of nightly observations of at least 40 bright nearby Sun-like stars \citep{Hall2018}. By combining the high precision of HARPS3 with the radical observation plan of THE, observers have the greatest chance at detecting low mass, long period exoplanets.
In fact, it has been shown that THE can detect Earth-twins in the habitable zone of Solar-type stars, for both single and multi-planet systems and with stellar signals \citep{Hall2018}.
Further, simulations have suggested that THE can outperform a typical reference survey and \textnormal{performs} comparably to an uninterrupted space-based schedule on accuracy of recovered parameters for Earth-twin detection \citep{Hall2018}.

\subsection{Summary}
In sum, there is currently little known about the planetary demographics of Earth-like and cool gas giant (CGG) exoplanets around Sun-like stars. As CGGs (Jupiter and Saturn) appear to have played a critical role in the formation, stability, and habitability of Earth, studying the occurrence rate of exoplanetary systems containing both an Earth-like exoplanet and a CGG-like exoplanet is a pressing question in understanding the frequency of Solar System analogs. \textnormal{Here we define a Solar System analog as a stellar system with a Sun-like star, a terrestrial planet with an Earth-like orbit, and a cool gas giant (CGG).} Additionally, as CGGs have larger signals than Earth-like exoplanets in several common observing methods, due to their larger mass and radius, determining a reasonable estimate for P(Earth $|$ CGG) would allow us to learn more about $\eta_\Earth$ (the mean number per star of rocky exoplanets with radii between 1 and 1.5–2 Earth-radii that reside in the optimistic habitable zone). There is a precedent for these types of Bayesian studies of exoplanetary demographics. \textnormal{THE will recover minimum mass estimates for any Earth-like planets and CGG-like planets around $\sim$40 nearby Sun-like stars.}

\textnormal{As noted in \citet{Winn2022}, combining astrometry with precision RV observations can be particularly useful as the high-precision RV surveys can detect planets with greater signal-to-noise than current astrometric surveys, but do not reveal the orbital inclination and thus only give a lower limit on the mass. Astrometric observations, even when lower signal-to-noise, can aid in resolving the orbital inclination, and thus breaking the mass-inclination degeneracy.}

\textnormal{In what follows, we investigate whether Gaia and Roman astrometry in the coming decade can serve this purpose. Specifically, we study what planetary parameters these missions will accurately recover and compare several Roman observing strategies in pursuit of these goals. In Section \ref{sec: methods} we present our simulation and modeling methods and in Section \ref{sec: results} we present and discuss the results of our simulations.}

\section{Methods} \label{sec: methods}
In brief, given a set of orbital parameters, a cadence of observations, a duration of observations, and assuming white noise with a fixed standard deviation set to the expected instrumental 1$\sigma$ measurement errors, we can create a simulated dataset of both radial velocity and astrometry observations for an exoplanet using \texttt{exoplanet}. We then model this simulated data, again using \texttt{exoplanet}, and determine whether these orbital parameters can be accurately recovered. We assumed a Solar mass star for all simulation in this work.

\subsection{\texttt{exoplanet} Code}
For both simulating the orbits of the exoplanets sampled in our paper and efficiently modeling the subsequent orbits, we use the \texttt{exoplanet} codebase. \texttt{exoplanet} is an open source toolkit that uses efficient gradient based sampling to model exoplanet data from transits, radial velocities, and/or astrometry \citep{Foreman-Mackey2021}. Practically, in its gradient-based inference, \texttt{exoplanet} uses methods such as Hamiltonian Monte-Carlo, No U-Turn Sampling, and variational inference, which can improve performance (especially when there are more than 10 parameters) by orders of magnitude vs. other modeling methods such as Markov-Chain Monte Carlo (MCMC) \citep{Foreman-Mackey2021}. In contrast to many other tools in the vast ecosystem of exoplanet and time domain modeling software, \texttt{exoplanet} is designed as a framework with which a pipeline can be generated -- rather than a finished code base with a ``push go'' button \citep{Foreman-Mackey2021}. Therefore, we used \texttt{exoplanet} as a base package, but developed our own pipeline, which is shared with the community as a public GitHub repository.\footnote{\url{https://github.com/dyahalomi/rv_and_astrometry}} \texttt{exoplanet} uses PyMC3, a model building language and inference engine that scales well in the many parameter regime \citep{Salvatier2016}.

\subsection{Simulated Parameters} \label{subsec: sim_params}

We wanted to develop an understanding of how a Roman astrometric observing program can best contribute to the detection of both Earth-like and CGGs around the THE stars that will already have a decade of Gaia precision astrometry. In order to do so, we simulated two-planet systems, each with an Earth and a CGG around a Solar mass star. In the first iteration we simulated a Jupiter-like CGG and in the second iteration we simulated a Saturn-like CGG. Specifically, by Earth-like, Jupiter-like, and Saturn-like we mean that we adopted the physical and orbital parameters of these planets in our solar system. These assumed planetary parameters can be seen in Table \ref{table: simulated_parameters}. We varied the astrometric precision of Roman (5, 10, and 20 [$\mu$as]), as white noise on top of the astrometric values, and the duration of the Roman astrometric survey (0, 5, and 10 [years]). \textnormal{Assuming a random distribution of orbital inclinations in the sky, the orbital configurations will be uniformly distributed along $\cos{i}$, and so we simulate the median value of this distribution, 60$^\circ$, for the Earth-like planet inclination.} We assumed a mutual inclinations consistent with those in our solar system (1.3 degrees for Jupiter and 2.5 degrees for Saturn). 

\textnormal{Over the course of the primary, 5-years, Gaia mission, there will be 68.9 field-of-view passages for a sky averaged ecliptic latitude (including a mission-level 20$\%$ ``dead-time'') \citep{Perryman2014}. Extrapolating this out to 10-years, we simulate 138 observations evenly spaced throughout the Gaia 10-year mission (or an observation every $\sim$26 days). As the stars in the THE catalog will be bright nearby Sun-like stars, we assume that they will have G magnitude less than 12 and thus adopt an along-scan accuracy per field-of-view passage for Gaia of $\sigma_\textrm{fov}$ = 34.2 $\mu$as \citep{Perryman2014}. We note that we are not making use of the actual Gaia scanning law observations of a real system, but rather assuming a uniform time-series for simplicity. In order to account for Gaia's scanning pattern, we assume a random scanning direction, $\psi$ along a uniform distribution between 0 and 90 degrees for each observation. As Gaia precision will be $\sim$5 times worse across-scan (AC) than along-scan (AL) direction \citep{Lindegren2012, Abbas2017}, we determine the Gaia simulated white noise standard deviations for RA and declination ($\sigma_\textrm{RA}$ and $\sigma_\textrm{dec}$) following:}

\begin{equation}
    \sigma_\textrm{RA} = \sigma_\textrm{RA, AL} + \sigma_\textrm{RA, AC}
\end{equation}
where,
\begin{equation}
    \sigma_\textrm{RA, AL} = \sigma_\textrm{fov} \sin(\psi)
\end{equation}
\begin{equation}
    \sigma_\textrm{RA, AC} = 5 \sigma_\textrm{fov} \cos(\psi)
\end{equation}
and 
\begin{equation}
    \sigma_\textrm{dec} = \sigma_\textrm{dec, AL} + \sigma_\textrm{dec, AC}
\end{equation}
where,
\begin{equation}
    \sigma_\textrm{dec, AL} = \sigma_\textrm{fov} \cos(\psi)
\end{equation}
\begin{equation}
    \sigma_\textrm{dec, AC} = 5 \sigma_\textrm{fov} \sin(\psi). \nonumber
\end{equation}

For the radial velocity data, we adopted THE's nightly cadence and 10-year duration, assuming 180 days of observations per year, with data gaps when the star is not visible, as described in \citet{Hall2018}. We also adopted the same white noise used to represent photon shot noise of 0.3 m s$^{-1}$ as presented in \citet{Hall2018}. \textnormal{We note that we do not model stellar activity effects in our simulations, which in practice will likely complicate the THE modeling. As a year-long orbit cannot be easily detected from the ground, we follow the precedent in \citet{Hall2018} and simulate Earth with an orbital period of 293 days.}

In order to account for the biases present in small number sampling and the randomness in the small number of white noise variations around the true RV and astrometric values that we assume, we ran 10 different simulations for both two-planet systems, each with a different realization of the white noise. For a complete set of simulated parameter values, see Table \ref{table: simulated_parameters} and Table \ref{table: observing_parameters}.

\begin{table} 
\centering
\caption{Simulated Parameter Values}
\label{table: simulated_parameters}
\begin{tabular}{ c  c }

\textbf{Stellar Parameters} & \\
\hline
\hline
Parameter & Values\\
\hline
M & 1 M$_\Sun$ \\
distance & 10 pc \\

& \\

\textbf{Earth Parameters} & \\
\hline
\hline
Parameter & Values\\
\hline
P & 293 days \\
M & 1 M$_\Earth$ \\
e & 0.0167 \\
i & 60 deg \\
$\omega$ & 102.9 deg \\
$\Omega$ & 0 deg \\
T$_\textrm{per}$ & 100 days \\
\hline
K & 0.08 m/s \\
$\Theta$ & 0.26 $\mu$as \\ 

& \\

\textbf{Jupiter Parameters} & \\
\hline
\hline
Parameter &  Values\\
\hline
P & 4327 days \\
M & 317.96 M$_\Earth$ \\
e & 0.0484 \\
i & 61.3 deg \\
$\omega$ & -85.7 deg \\
$\Omega$ & 100.4 deg \\
T$_\textrm{per}$ & 500 days \\
\hline
K & 10.95 m/s \\
$\Theta$ & 497.02 $\mu$as \\

& \\

\textbf{Saturn Parameters} & \\
\hline
\hline
Parameter &  Values\\
\hline
P & 10740 days \\
M & 95.16 M$_\Earth$ \\
e & 0.0539 \\
i & 62.5 deg \\
$\omega$ & -21.1 deg \\
$\Omega$ & 113.7 deg \\
T$_\textrm{per}$ & 970 days \\
\hline
K & 2.45 m/s \\
$\Theta$ & 272.70 $\mu$as \\ 

\end{tabular}
\end{table}

\begin{table} 
\centering
\caption{Observing Parameters}
\label{table: observing_parameters}
\begin{tabular}{ c  c  c }

Parameter & Varied & Values\\
\hline
THE White Noise 1$\sigma$ &  & 0.3 ms$^{-1}$ \\
Roman White Noise 1$\sigma$ & X & [5, 10, 20] $\mu$as \\
Gaia AL White Noise 1$\sigma$ &  & 34.2 $\mu$as \\
Gaia AC White Noise 1$\sigma$ &  & 171 $\mu$as \\
\hline

THE Duration &  & 10 years \\
Roman Duration & X & [0, 5, 10] years \\
Gaia Duration &  & 10 years \\
\hline

THE Cadence &  & 1 per day \\
Roman Cadence &  & 1 per 40 days \\
Gaia Cadence &  & 1 per 26 days \\
\hline
Gaia Scan Angle & X & $\mathcal{U}(0^\circ, 90^\circ$) \\

\hline
\end{tabular}
\end{table}

Astrometric data is often reported and modeled in position angle and separation over time. However, position angle doesn't have an independent uncertainty since it depends on the amplitude of the separation. Therefore, for the Gaia and Roman data, we simulated and modeled right ascension and declination as it is more straight forward to apply white noise to the simulated data.

\subsection{Modeling Pipeline}
\subsubsection{Radial Velocity Model}
We assumed that THE RVs will provide a period with a precision of at worst 10$\%$, based on the simulations presented in \citet{Hall2018}, and initialize our model period using a uniform distribution centered at the true period with a lower-bound and upper-bound 10$\%$ off from the true period, and call this period value P$_0$. Then using the \texttt{exoplanet.estimate\_semi\_amplitude()} function, that finds the maximum likelihood semi-amplitude for a circular orbit at fixed period, we generate an initial guess for the semi-amplitude (K$_0$) using P$_0$ and the RV data. Next, we generate an initial guess for the RV parameters, using ``minimization via scheduling.'' In minimization via scheduling, one iteratively minimizes specific parameters one at a time, before minimizing over all parameters at once. This allows us to optimize parameters for which we know little about, before performing the more computationally expensive process of optimizing all model parameters simultaneously. While minimizing via scheduling can in some cases cause parameters to land in local (rather than global) minima, we found that this process consistently helped us generate a more accurate initial guess for the RV model and this method is recommended for \texttt{exoplanet} \citep{Foreman-Mackey2021}. We sample period (P), eccentricity (e), argument of the periapsis of the star ($\omega$), radial velocity semi-amplitude (K), and the time of periapsis passage (T$_\textrm{per}$). As is standard in modeling of exoplanet radial velocity data, we sampled the eccentricity and argument of periapsis as $\sqrt{\textrm{e}}\cos{\omega}$ and $\sqrt{\textrm{e}}\sin{\omega}$ \citep{Eastman2013, Foreman-Mackey2021}. Additionally, we sampled period and semi-amplitude in natural logarithmic space. The 5 radial velocity modeling parameters, priors, and starting points can be seen in Table \ref{table: rv_params}. 

\textnormal{We would like to highlight several assumptions that we have made in our simulated THE data and RV model. (1) We assume that there are no additional RV signals from other planets in the system, besides the Earth-like and CGG-like planets, which could further complicate the period determination. (2) We assume that any stellar activity induced RV variability are perfectly understood. (3) We assume that models of the THE-only dataset will recover the planetary periods within 10$\%$ of their true values. \citet{Hall2018} showed that multiple peaks may appear in the periodogram for Earth-like planets recovered with THE, which would thus require further analysis of which peak period is the ``true'' RV signal.}

\begin{table} 
\centering
\caption{Parameters per Planet for the RV MCMC Model}
\label{table: rv_params}
\begin{tabular}{ c  c  c }
\hline
Parameter & Prior\footnote{\label{fn: priorLC} $\mathcal{U}(x,y)$ denotes a uniform distribution between x and y.} & Starting Point\footnote{\label{fn: starting_point} ``...'' denotes no starting point used in minimization.} \\

\hline 
$\ln \textrm{P}$  & $\mathcal{U}$(0, 11) &  $\ln \textrm{P}_0$ \\
$\ln \textrm{K}$  & $\mathcal{U}$(-4, 4) &  $\ln \textrm{K}_0$  \\
$\sqrt{\textrm{e}}\cos{\omega}$ & kipping13\footnote{\label{kipping_foot} Beta distribution prior applied based on \citet{Kipping2013}} & 0.01  \\
$\sqrt{\textrm{e}}\sin{\omega}$ & kipping13$^{\ref{kipping_foot}}$ & 0.01  \\
T$_\textrm{per}$ & $\mathcal{U}$(0, $\textrm{P}_0$) & 0.5 $\textrm{P}_0$ \\

\hline
\end{tabular}
\end{table}

In modeling the simulated RV data, we use the standard Keplerian equations. Including the magnitude of barycentric motion of the star ($\gamma$), which we set to zero, the radial velocity signal is described by:

\begin{equation} \label{eq: RVsignal}
    RV(\nu) = K \Big[\cos \big(\nu(t) + \omega \big) + e \cos \omega \Big] + \gamma
\end{equation}

where

\begin{equation} \label{eq: rv_signal}
    K = (\frac{2\pi G}{P})^{\frac{1}{3}} \frac{\textrm{M}_{p} \sin{i}}{(\textrm{M}_{p}+\textrm{M}_{*})^\frac{2}{3}} \frac{1}{\sqrt{1-e^2}}.
\end{equation}

\subsubsection{Joint Radial Velocity and Astrometric Model}
Now that we have an initial guess for the exoplanetary parameters from the radial velocity minimization, we can include the astrometric data and jointly sample radial velocity and astrometric parameter space using both simulated datasets in order to recover a complete posterior sample. As we are testing different astrometric observational cadences and durations, modeling the RV observations independently first has the benefit of speeding up the joint modeling. By using the RV minimization as a starting point for all parameters besides the position angles of the ascending node ($\Omega$) and splitting the mass-inclination degeneracy (m$_\textrm{p}$ and i), we can more efficiently model the larger parameter space. We again minimize the joint RV and astrometry model via scheduling, to obtain an initial guess for the parameters in the model. In order to get an accurate first guess for mass and inclination, as RVs provide weak information about inclination, we initialize at 5 different inclinations test values (i$_\textrm{test}$) of 5, 25, 45, 65, and 85 degrees. We determine a corresponding mass test value (m$_\textrm{test}$), based on each i$_\textrm{test}$ value, by dividing the minimum mass from the radial velocity model by sin(i$_\textrm{test}$). We then minimize as described above for all 5 initial guesses, and pick the minimized MAP solution with the highest likelihood value. 

We found the position angles of the ascending nodes for the two exoplanets to be co-dependent in the joint astrometric and radial velocity posteriors. Therefore, for more efficient sampling, we model position angles of the ascending nodes as $\Omega_+$ and $\Omega_-$, where 

\begin{equation}
    \Omega_+ = \frac{\Omega_1 + \, \Omega_2}{2} \ \ \ \ \ \textrm{and} \ \ \ \ \
    \Omega_- = \frac{\Omega_1 - \, \Omega_2}{2}.
\end{equation}

\textnormal{As mass and inclination are highly correlated, due to the mass-inclination degeneracy in RV observations, we model mass and inclination as $\sqrt{\textrm{m}_\textrm{p}} \sin{\textrm{i}}$ and $\sqrt{\textrm{m}_\textrm{p}} \cos{\textrm{i}}$. We assume that Gaia will provide a precise and accurate constraint on the parallax, and so sample the parallax with a prior equal to a normal distribution centered on the true parallax, 0.1 as, with a standard deviation of 0.01 as.}

The complete set of modeled parameters, priors, and starting points for the joint radial velocity and astrometric model can be seen in Table \ref{table: joint_params}.

For both the Earth-Jupiter and Earth-Saturn systems, we simulated and modeled 10 random realizations of white noise scatter for each observing strategy in order to account for biases in the small number sampling of the observations. For each simulated dataset, we run an MCMC with \texttt{exoplanet}, using 2 chains, 4000 tuning steps, and 4000 drawing steps. We checked the MCMC posteriors for convergence by only keeping those for which the $\hat{R}$ statistic for the masses and periods were less than 1.05.

\begin{table} 
\centering
\caption{Parameters per Planet for the Joint RV and Astrometric MCMC Model}
\label{table: joint_params}
\begin{tabular}{ c  c  c }
\hline
Parameter & Prior\footnote{\label{fn: priorLC} Priors adopted in the MCMC model. $\mathcal{U}(x,y)$ denotes a uniform distribution between x and y. $\mathcal{N}(\mu, \sigma^2$) denotes a Gaussian distribution centered at $\mu$ with a standard deviation of $\sigma$.} & Starting Point\footnote{\label{fn: starting_point} ``...'' denotes no starting point used in minimization.}\\
\hline 

$\ln \textrm{P}$  & $\mathcal{U}$(0, $\log 2\textrm{P}_\textrm{RV}$) &  $\ln \textrm{P}_\textrm{RV}$ \\

$\sqrt{\textrm{e}}\cos{\omega}$ & kipping13\footnote{\label{kipping_foot2} Beta distribution prior applied based on \citet{Kipping2013}} & $\sqrt{\textrm{e}}\cos{\omega}_\textrm{RV}$  \\

$\sqrt{\textrm{e}}\sin{\omega}$ & kipping13$^{\ref{kipping_foot2}}$ & $\sqrt{\textrm{e}}\sin{\omega}_\textrm{RV}$  \\

phase & $\mathcal{U}$(-$\pi$, -$\pi$) & phase$_\textrm{RV}$  \\

$\sqrt{\textrm{m}_\textrm{p}} \sin{\textrm{i}}$ & $\mathcal{U}$(0, 10$\sqrt{\textrm{m}_\textrm{test}} \sin{\textrm{i}_\textrm{test}}$)  & $\sqrt{\textrm{m}_\textrm{test}} \sin{\textrm{i}_\textrm{test}}$  \\

$\sqrt{\textrm{m}_\textrm{p}} \cos{\textrm{i}}$ & $\mathcal{U}$(0, 10$\sqrt{\textrm{m}_\textrm{test}} \cos{\textrm{i}_\textrm{test}}$)  & $\sqrt{\textrm{m}_\textrm{test}} \cos{\textrm{i}_\textrm{test}}$  \\

$\Omega_{+}$ & $\mathcal{U}$(-$\pi$, $\pi$) & 0 \\
$\Omega_{-}$ & $\mathcal{U}$(-$\pi$, $\pi$) & 0 \\

$\pi$  & $\mathcal{N}$(0.1, $0.01^2$) &  0.1 \\
\hline
\end{tabular}
\end{table}

In modeling the astrometric data, we used the following astrometric equations, which are built into \texttt{exoplanet} and explained in \textnormal{detail} in \citet{Quirrenbach2010}. Including both parallax and proper motion, the total astrometric signal is described by $\xi$ and $\eta$, the standard or projected coordinates. These coordinates are the projection of RA and Dec onto the tangent plane of the sky, and are defined below:

\begin{equation} \label{eq: astrometric_xi}
   \xi (t) = \alpha_0^\star + \pi_{\alpha^\star}{\pi} + (t - t_0)\mu_{\alpha^\star} + B \, X(t) + G \, Y(t)
\end{equation}
\begin{equation}
    \eta (t) = \delta_0 + \pi_{\delta}{\pi} + (t - t_0)\mu_{\delta} + A \, X(t) + F \, Y(t)
\end{equation}    

where $\alpha_{\star} = \alpha \cos \delta$, $(\alpha_0^\star, \delta_0)$ is the reference right ascension and declination of the star, ($\pi_{\alpha^\star}, \pi_{\delta}$) is the parallax factors describing the parallactic ellipse of the star, and ($\mu_{\alpha^\star}, \mu_{\delta}$) is the proper motion vector of the star. The equation above is parametrized by the Thiele-Innes constants:

\begin{equation}
    A = \Theta (\cos\Omega \cos\omega - \sin\Omega \sin\omega \cos i)
\end{equation}
\begin{equation}
    B = \Theta (\sin\Omega \cos\omega + \cos\Omega \sin\omega \cos i)
\end{equation}   
\begin{equation}
    F = \Theta (-\cos\Omega \sin\omega - \sin\Omega \cos\omega \cos i)
\end{equation}    
\begin{equation}
    G = \Theta (-\sin\Omega \sin\omega + \cos\Omega \cos\omega \cos i)
\end{equation}

where $\Theta$ is the semi-major axis of the apparent ellipse or what we will call the astrometric signal due to an orbiting exoplanet, in angular units. The semi-major axis of the apparent ellipse can be solved using Kepler's Third Law (where m$_{\textrm{tot}}$ is the total mass, or for a single exoplanet system $\textrm{m}_{\textrm{tot}} = \textrm{m}_{\star} + \textrm{m}_{\textrm{p}}$):

\begin{equation} \label{eq: astrometric_signal}
\Theta = [\frac{P^2 \: G \: \textrm{m}_{\textrm{tot}}}{4\pi^2}]^{\frac{1}{3}}
\end{equation}

If we assume a circular orbit and that $\textrm{m}_\textrm{p} \ll \textrm{m}_{\star}$. At a distance ``d'' = $\frac{1}{\pi}$ from the observer and an orbital radius ``R'' = a (where R and d must be in the same units), the astrometric signal equals:

\begin{equation}
\Theta = 
\frac{\textrm{m}_\textrm{p}}{\textrm{m}_{\star}} \frac{\textrm{R}}{\textrm{d}} = 
\Big(\frac{\textrm{G}}{4\pi^2} \Big)^{1/3}  
\frac{\textrm{m}_\textrm{p}}{\textrm{m}_{\star}^{2/3}}
\frac{\textrm{P}^{2/3}}{\textrm{d}}
\end{equation}
\begin{equation}
\Theta = 
3 \, \mu \textrm{as} \cdot
\Big(\frac{\textrm{m}_\textrm{p}}{\textrm{M}_{\bigoplus}} \Big)
\Big(\frac{\textrm{m}_\star}{\textrm{M}_{\bigodot}} \Big)^{-2/3}
\Big(\frac{\textrm{P}}{\textrm{yr}} \Big)^{2/3}
\Big(\frac{\textrm{d}}{\textrm{pc}} \Big)^{-1}
\end{equation}

Lastly, X(t) and Y(t), dimensionless auxiliary values often referred to as the ``normalized coordinates'', are the vectors describing the motion of the star caused by its exoplanetary orbit. They are defined by:

\begin{equation}
X(t) = cos(E(t)) - e
\end{equation}
\begin{equation}
Y(t) = \sqrt{1-e^2} \sin(E(t))
\end{equation}

where E is the eccentric anomaly, and the solution of Kepler's equation:
\begin{equation}
E = \frac{2\pi}{P} (t - T) + e\sin E
\end{equation}

\textnormal{Gaia will not provide RA and DEC observations, but rather time-series observations in a local coordinate system projected along the direction of the scanning of the telescope. For a detailed description of the Gaia astrometric observations, see \citet{Holl2023}. In short, Gaia will provide an along-scan observable:}
\begin{equation}
    w^\textrm{(model)} = w_\textrm{ss} + w_\textrm{k1}
\end{equation}
where, $w_\textrm{ss}$ is the single-source model and $w_\textrm{k1}$ is the single Keplerian model:
\begin{equation}
    w_\textrm{ss} = (\Delta\alpha^* + \mu_{\alpha^*}\, t)\sin\psi + (\Delta\delta + \mu_{\delta}\, t)\cos\psi + \varpi \, f_\varpi, 
\end{equation}
\begin{equation}
    w_\textrm{k1} = (BX + GY)\sin\psi + (AX + FY)\cos\psi.
\end{equation}

\textnormal{Here $\Delta\alpha^*$ = $\Delta\alpha\cos\delta$ and $\Delta\delta$ are small offsets in equatorial coordinates from some fixed reference point ($\alpha_0$, $\delta_0$), $\mu_{\alpha^*}$ and $\mu_{\delta}$ are proper motions in these coordinates, t is the time since reference time J2016.0, $\varpi$ is the parallax, $f_\varpi$ is the parallax factor, and $\psi$ is the scan angle (which is defined such that $\psi$ = 0 when the field-of-view is moving towards local North and $\psi$ = 90$^\circ$ towards local East\footnote{\url{https://www.cosmos.esa.int/web/gaia/scanning-law-pointings}}).}

\textnormal{In order to account for this change, in simulating our Gaia astrometry we assumed a random scan angle and converted from the along-scan and across-scan frame to the RA and DEC frame, carrying the factor of $\sim$5 times larger error in the across-scan direction \citep{Lindegren2012, Abbas2017}. For more information, see Section \ref{subsec: sim_params}.}

Lastly, we highlight that the synergistic nature of the astrometric signal and the radial velocity signal, which was previously shown in Figure \ref{fig: signal_strength}, can also be understood mathematically by looking at the scaling of radial velocity semi-amplitude with period in Equation \ref{eq: rv_signal} and the scaling of semi-major axis with period in Equation \ref{eq: astrometric_signal}. Here we see that the astrometric signal varies with period$^{2/3}$ while the radial velocity signal varies with period$^{-1/3}$.

\textnormal{In our simulations, we assume stellar mass is known with infinite precision. This in practice will obviously not be the case, but as we do not yet know the stars in the THE catalog or the expected stellar mass precision of those stars, we decided to not add uncertainties based on stellar mass determination in our simulations or the discussion that follows. Current asteroseismic fits to Kepler observations of main-sequence stars provide masses with $\sim$4$\%$ uncertainties \citep{SilvaAguirre2017}. If asteroseismic data isn't available, one might expect stellar mass determination with uncertainties of order $\sim$10$\%$ \citep[ie. TESS Input Catalog:][]{Stassun2019}. These uncertainties would clearly inflate the following reported precisions. As such, one could consider the reported planetary mass precisions as a planet-to-star mass-ratio precisions.}

\section{Results} \label{sec: results}

\begin{figure*}[htb]
\centering 
\includegraphics[width=\textwidth]{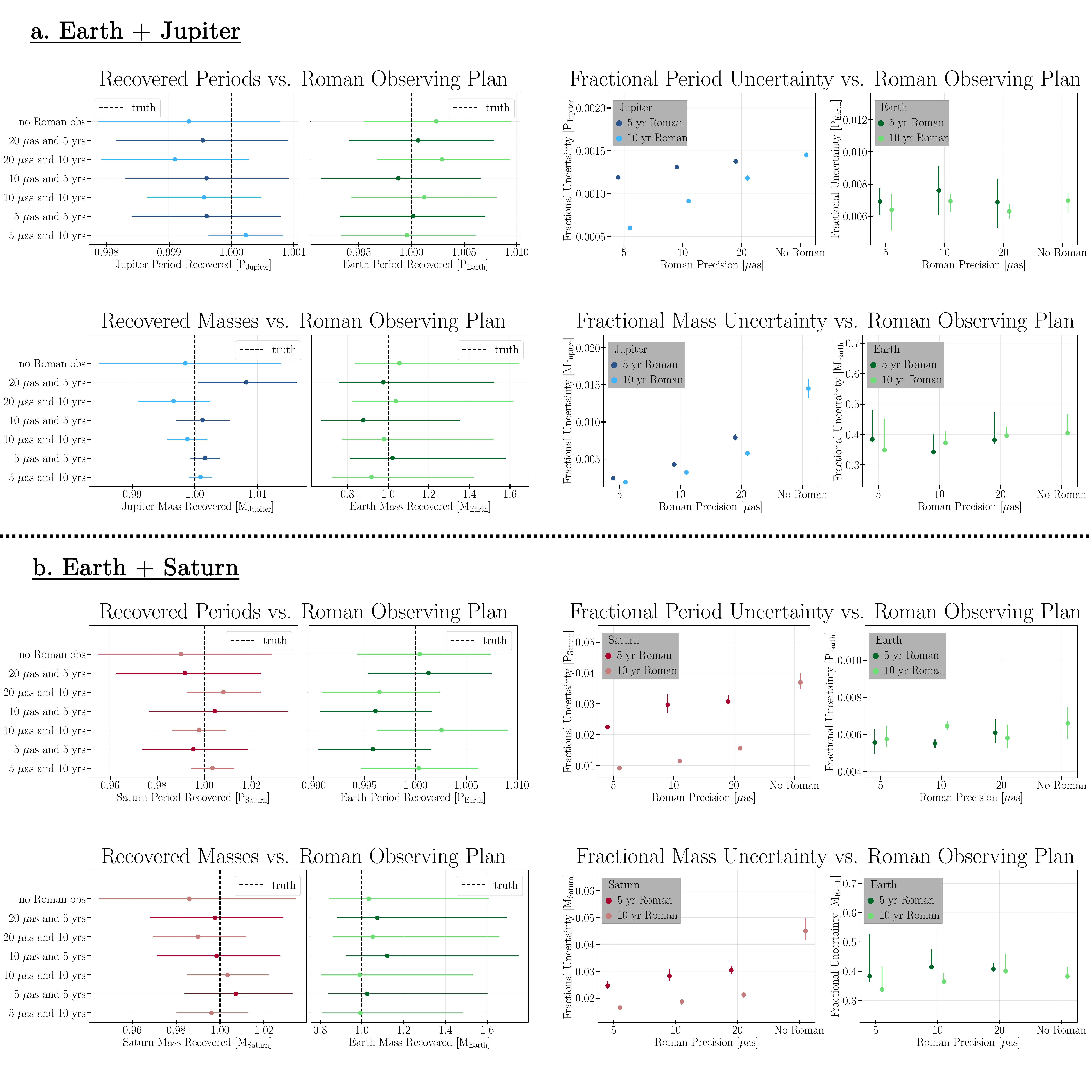}
\caption{$\textbf{Left:}$ Recovered median and 1$\sigma$ MCMC period and mass vs. Roman observing plan for exoplanetary systems with an Earth-like exoplanet and CGG-like companion exoplanet (a: Jupiter-like, b: Saturn-like) around a Sun-like star at 10 parsecs. Each data point represents the median of the multiple iterations of simulations that we ran and subsequent MCMC models that we fit in order to account for correlated noise and biases in our small number of astrometric observations. The error bars are the median of the MCMC 1$\sigma$ results. $\textbf{Right:}$ Period and mass 1$\sigma$ uncertainty vs. Roman observing plan. Here, the data points are the mean of the upper and lower 1$\sigma$ error bars on the left plots. The error bars on these plots represent the median absolute deviation amongst the 1$\sigma$ values for the different MCMC runs. $\textbf{Takeaway:}$ These figures show 3 main things: (1) An observing program with only THE RVs and Gaia astrometry can recover Earth-like and CGG exoplanets around nearby Sun-like stars. (2) A modest astrometric observing program with Roman of the THE stars could extend the astrometric baseline to $\sim$25 years and improve the precision of the recovered CGG masses and periods. (3) Roman astrometry will not improve the precision of the detection of Earth-like exoplanets around stars at 10 parsecs beyond the Gaia astrometric observations.}
\label{fig: main_results}

\end{figure*}

\subsection{A Survey with only THE and Gaia Data}

In order to evaluate precision and accuracy of the different observing strategies we focus on period and mass for both exoplanets in the simulated exoplanetary system. In short, a survey of Sun-like stars at 10 parsecs, with only THE RVs and Gaia astrometry could detect Earth-like within $\sim$50$\%$ of the true mass and within $\sim$1$\%$ of the true period, Jupiter-like within $\sim$2$\%$ of the true mass and within $\sim$0.2$\%$ of the true period, and Saturn-like within $\sim$5$\%$ of the true mass and within $\sim$4$\%$ of the true period. This can be seen in Figure \ref{fig: main_results} as the masses and periods for both the Earth-like and CGG-like exoplanets are recovered. This can also be seen in Table \ref{tab: earth_jup_period}, Table \ref{tab: earth_jup_mass}, Table \ref{tab: earth_sat_period}, and Table \ref{tab: earth_sat_mass}, which show the recovered median and 1$\sigma$ periods and masses for the Earth-like and CGG-like exoplanets.

\subsection{Precision Improvement from Roman Astrometry}
Including Roman observations improves the precision of the mass and period determination of the CGGs. \textnormal{An optimal Roman observing program (5 $\mu$as precision and 10 years of observations) could detect Jupiter analogs within $\sim$0.2$\%$ of the true mass and within $\sim$0.06$\%$ of the true period and Saturn analogs within $\sim$2$\%$ of the true mass and within $\sim$0.9$\%$ of the true period. For CGGs, therefore, adding Roman astrometry will improve the precision of mass detections by a factor up to $\sim$10 and the precision of the period detections by a factor up to $\sim$4.} 

For Earth analogs, additional Roman observations won't improve the detection precision on top of Gaia observations. This can be seen in Figure \ref{fig: main_results} as the mass and period uncertainties for the CGG-like exoplanets are positively correlated with the Roman precision and duration, but for Earth-like exoplanets there is no correlation. \textnormal{This is likely because for Earth-like exoplanets around Sun-like stars at 10pc, Gaia astrometry is sufficient to apply loose constraints on the inclination of the orbit. Roman astrometry is unable to significantly improve the inclination determination, as the astrometric signal is still too low for the assumed observing plan to fully resolve the astrometric orbit with high signal-to-noise. As the inclination is similarly loosely constrained by both Gaia and Roman for the Earth-like planet, the mass is also only loosely constrained ($\sim$50$\%$).} This can also be seen in Table \ref{tab: earth_jup_period}, Table \ref{tab: earth_jup_mass}, Table \ref{tab: earth_sat_period}, and Table \ref{tab: earth_sat_mass}, which show the recovered median and 1$\sigma$ periods and masses for the Earth-like and CGG-like exoplanets.

\begin{figure*}[!htb]
\centering 
\includegraphics[width=\textwidth]{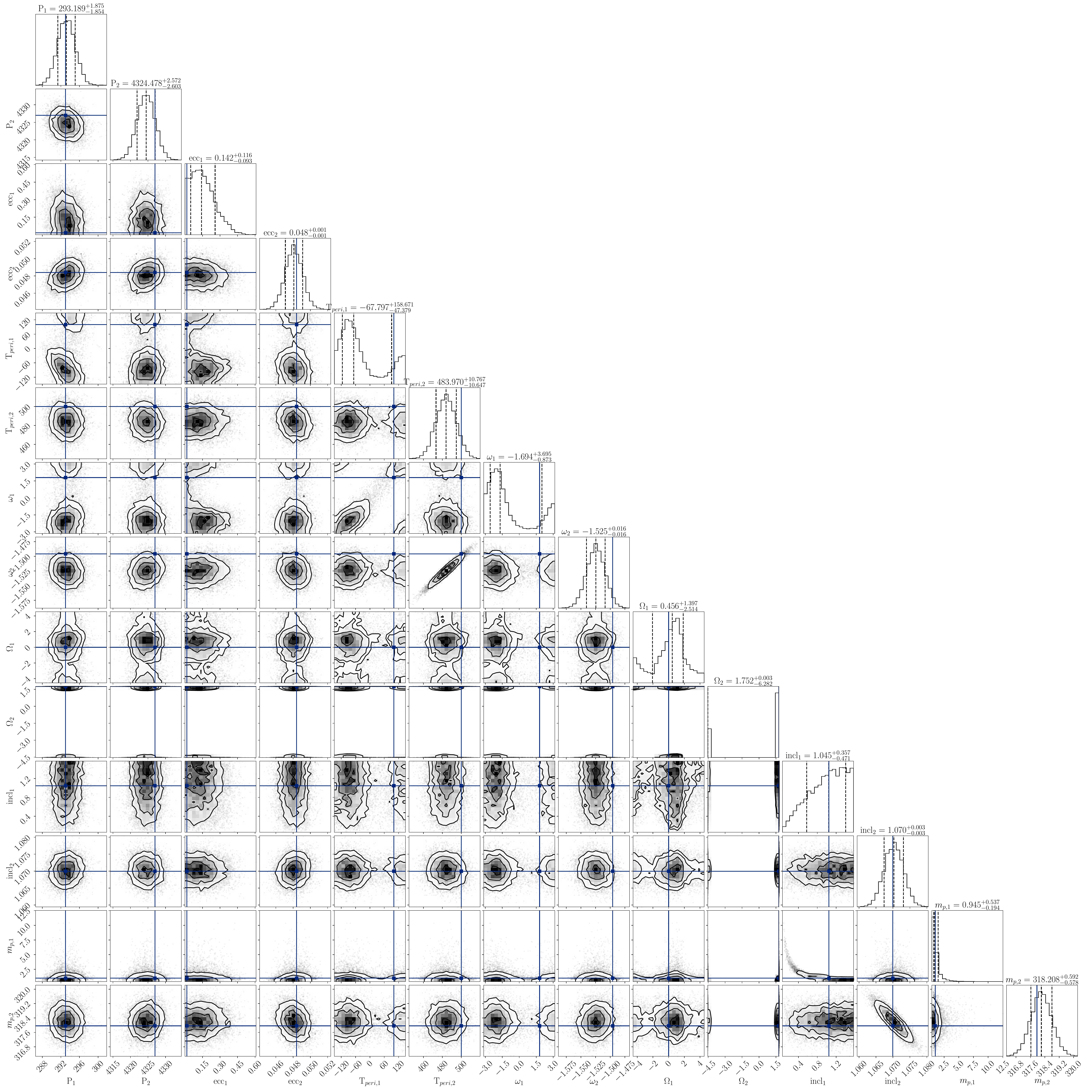}
\caption{Corner plot for Roman observing program with 10 years of  observations every 40 days and an assumed 1$\sigma$ measurement error of 5 $\mu$as. Simulated data is of a Jupiter-like exoplanet (subscript 1 above) and an Earth-like exoplanet 
(subscript 2 above) orbiting a Sun-like star at 10 parsecs. Plot generated with \texttt{corner}.}
\label{fig: corner}

\end{figure*}

The fact that Roman improves the detection precision for CGGs but not Earth analogs can be understood by considering the sensitivity of RV and astrometry observations and the observing durations and cadences of the three observing programs. Looking at Figure \ref{fig: signal_strength}, we can see that even with optimal Roman 1$\sigma$ measurement error of 5$\mu$as, the astrometric signal of an Earth analog at 10 parsecs still lies well below the 1$\sigma$ measurement error. This is mitigated by the large number of precision RV observations. Simply put Earth analog detection from a joint RV (THE) and astrometry (Gaia + Roman) program is dominated by the THE's expected $\sim$1800 precision RV observations per target over 10 years. For CGGs, with astrometric signals that are easily detectable by Gaia and Roman's expected astrometric precision (evident by the precise determinations of CGG inclinations), including additional Roman astrometry will extend the baseline of astrometric observations to several decades from Gaia's start in 2013 through a possible Roman observing program ending in the late 2020s. In so doing, this $\sim$25 year astrometric baseline will provide tighter constraints on the CGG planetary and orbital parameters, allowing for very accurate models of the exoplanets' physical and orbital parameters. Said differently, as previously stated, astrometric models are highly degenerate without a baseline of observation at least as long as the orbital period \citep{Casertano2008} -- so for CGGs with greater than decade long orbits, extending the astrometric baseline, up to potentially $\sim$25 years with Roman, not surprisingly improves the detection precision.

\subsection{Discussion} \label{discussion}

Obtaining precise orbital periods for potential CGGs outside Earth analogs will be essential in studying the near mean motion resonance effect found for multi-planet systems in closer in orbits \citep{Petrovich2013}. Therefore, the addition of several well timed Roman observations per year throughout its lifetime, which could allow for 1$\%$ detections of CGG periods out to Saturn-like orbits, would be crucial to understanding the orbits of exoplanetary systems that resemble our Solar System.

The MCMC posteriors for all parameters, from a single simulation of a Roman observing program with 5$\mu$as precision and 10 years of observations can be seen in the form of a corner plot in Figure \ref{fig: corner}. The true values of the simulated parameters are also shown in Figure \ref{fig: corner} as the blue lines in each plot. The parameter space appears to be well sampled and most parameters are not tightly correlated. The position angles of the ascending nodes ($\Omega_1$ and $\Omega_2$) stand out as they are quite poorly constrained. Another thing of note, is that the eccentricity of the Earth-like exoplanet is not very well constrained. This may be because the eccentricity and argument of periastron ($\omega$) posteriors are correlated and the argument of periastron is completely degenerate with the time of periastron. As the Earth's eccentricity is poorly constrained, we found it very helpful in expediting the fit to include a prior on the eccentricities from \citet{Kipping2013}. Finally, we find that the masses and inclinations are correlated. \textnormal{As the Earth-like inclinations are poorly constrained due to astrometric precision limitations, the Earth-like planet masses are only constrained to 50$\%$ whether using only Gaia astrometry or both Gaia and Roman astrometry.}

Further, closely analyzing the correlation of the modeled parameters in a joint radial velocity and astrometric survey is an important topic as we approach the era of astrometric exoplanet detections with Gaia and Roman. Further, it would be interesting to investigate the covariance and co-dependence of the multiple exoplanet models in joint astrometric and radial velocity surveys. Specifically, one could study the impact of earth detection with and without cool gas giants and or other exoplanets in the system.

\textnormal{For further reading on the combination of Gaia astrometry and radial velocity observations, we point the reader to \citet{DelisleSegransan2022} and \citet{Sozzetti2023}. \citet{DelisleSegransan2022} presents a Fourier analysis of the orbital elements of Gaia astrometry in combination with radial velocities. \citet{Sozzetti2023} presents an example of combining Gaia DR3 astrometry with archival and intensive follow-up RV monitoring to precisely model an exoplanetary system and is the first exoplanet candidate astrometrically detected by Gaia to be successfully confirmed based on RV follow-up observations.}

\section{Conclusion}

We have simulated and modeled a joint radial velocity (Terra Hunting Experiment) and astrometric (Gaia + Roman) observing program, which is capable of recovering Earth-like, Jupiter-like, and Saturn-like exoplanets around bright nearby Sun-like stars. \textnormal{Assuming stellar mass is known with infinite precision}, we show that a program with only THE and Gaia observations could detect with precisions of ($\sim$50$\%$, $\sim$2$\%$, $\sim$5$\%$) of the true mass and with precisions of ($\sim$1$\%$, $\sim$0.2$\%$, $\sim$4$\%$) of the true period for Earth, Jupiter, and Saturn-like exoplanets respectively. Further, we show that adding in a modest targeted observing program with Roman could improve the precision of detections to ($\sim$0.2$\%$, $\sim$2$\%$) of the true mass, an improvement by as much as a factor of $\sim$10, and to ($\sim$0.06$\%$, $\sim$0.9$\%$) of the true period, an improvement by as much as a factor of $\sim$4, for Jupiter and Saturn analogs, respectively. Obtaining precise masses and periods for Earth, Jupiter, and Saturn analogous exoplanets will profoundly impact our understanding of the occurrence rate of Solar System analogs -- a result that has previously eluded our grasp due to instrumental limitations. 

While we did not simulate three planet systems, we believe that the longer ($\sim$25 years) baseline will be particularly important for resolving three planet systems
(Earth-Jupiter-Saturn). For these three planet systems, the longer astrometric baseline will likely be essential for detecting Saturn analogs and obtaining an accurate Jupiter analog mass. Future work should further investigate these three planet systems.

\begin{acknowledgments}
\textnormal{DAY thanks the LSSTC Data Science Fellowship Program, which is funded by LSSTC, NSF Cybertraining Grant \#1829740, the Brinson Foundation, and the Moore Foundation; his participation in the program has benefited this work.}

We thank Megan Bedell, Adrian Price-Whelan, and David Kipping for help and guidance throughout the project. We also would like to thank members of the Astronomical Data Group at the Flatiron Institute Center for Computational Astronomy's and the Columbia Astronomy Department for fruitful discussions.

This research made use of \textsf{exoplanet} \citep{exoplanet:joss, exoplanet:zenodo} and its dependencies \citep{exoplanet:arviz, exoplanet:astropy13, exoplanet:astropy18, exoplanet:kipping13b, exoplanet:pymc3, exoplanet:theano}.
\end{acknowledgments}

%


\software{
\texttt{exoplanet} \citep{Foreman-Mackey2021}, $\,$
\texttt{matplotlib} \citep{matplotlib}, $\,$
\texttt{numpy} \citep{numpy}, $\,$
\texttt{scipy} \citep{scipy}$\,$
\texttt{PyMC3} \citep{Salvatier2016} $\,$
\texttt{corner} \citep{Foreman-Mackey2016}.}



\bibliography{main}{}
\bibliographystyle{aasjournal}

\newpage
\appendix

Below are tables showing the recovered MCMC median and 1$\sigma$ results for the Earth-like and CGG-like exoplanets around Sun-like stars at 10 parsecs. Each median value in these tables represents the median of the multiple iterations of simulations that we ran and subsequent MCMC models that we fit in order to account for correlated noise and biases in our small number of astrometric observations. The error bars are the median of the MCMC 1$\sigma$ results.

\begin{table}[!htb]
\centering 
\includegraphics[width=\textwidth]{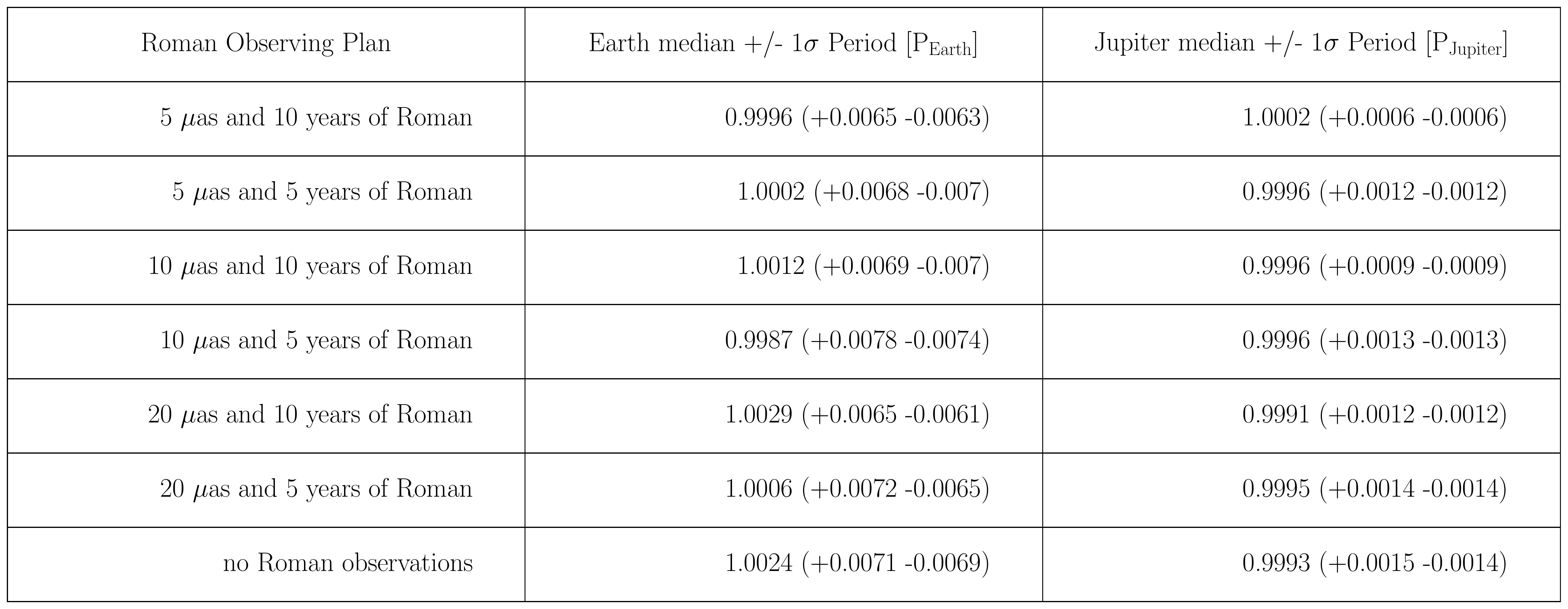}
\caption{Recovered MCMC median and 1$\sigma$ periods for the simulated Earth and Jupiter around a Sun at 10 parsecs.}
\label{tab: earth_jup_period}

\end{table}

\begin{table}[!htb]
\centering 
\includegraphics[width=\textwidth]{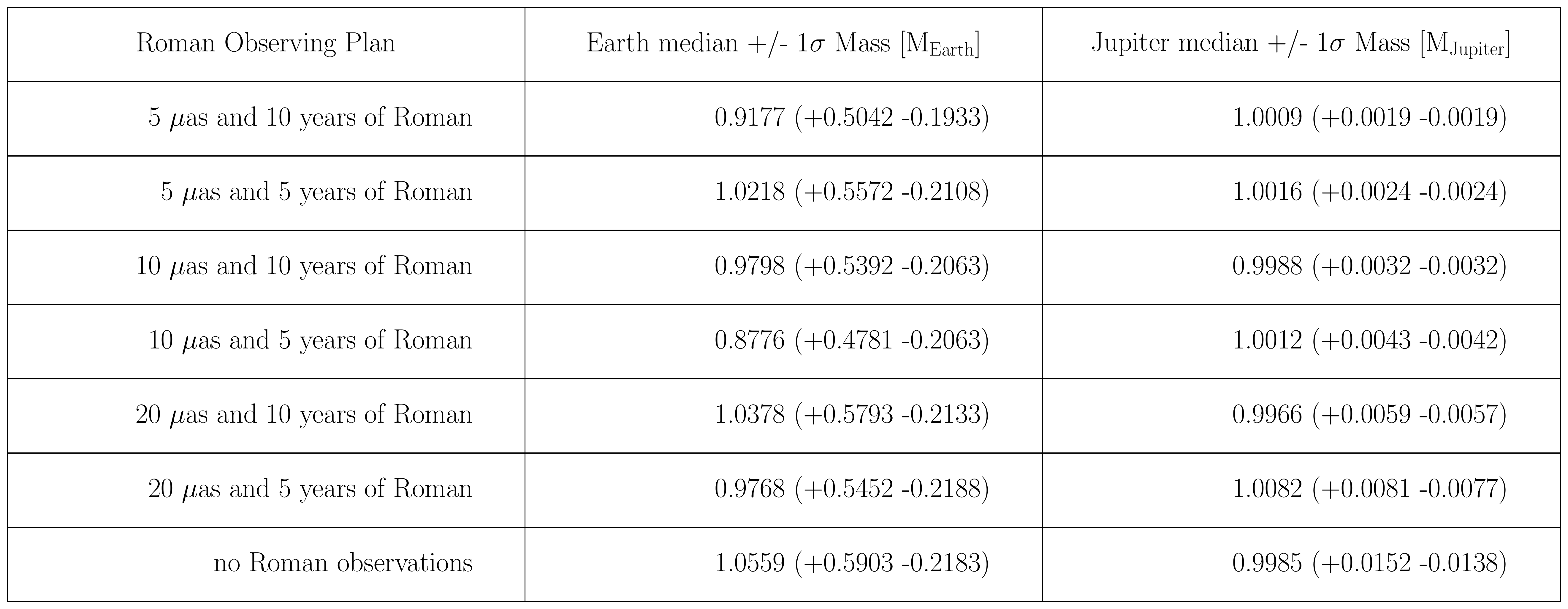}
\caption{Recovered MCMC median and 1$\sigma$ periods for the simulated Earth and Jupiter around a Sun at 10 parsecs.}
\label{tab: earth_jup_mass}

\end{table}

\begin{table}[!htb]
\centering 
\includegraphics[width=\textwidth]{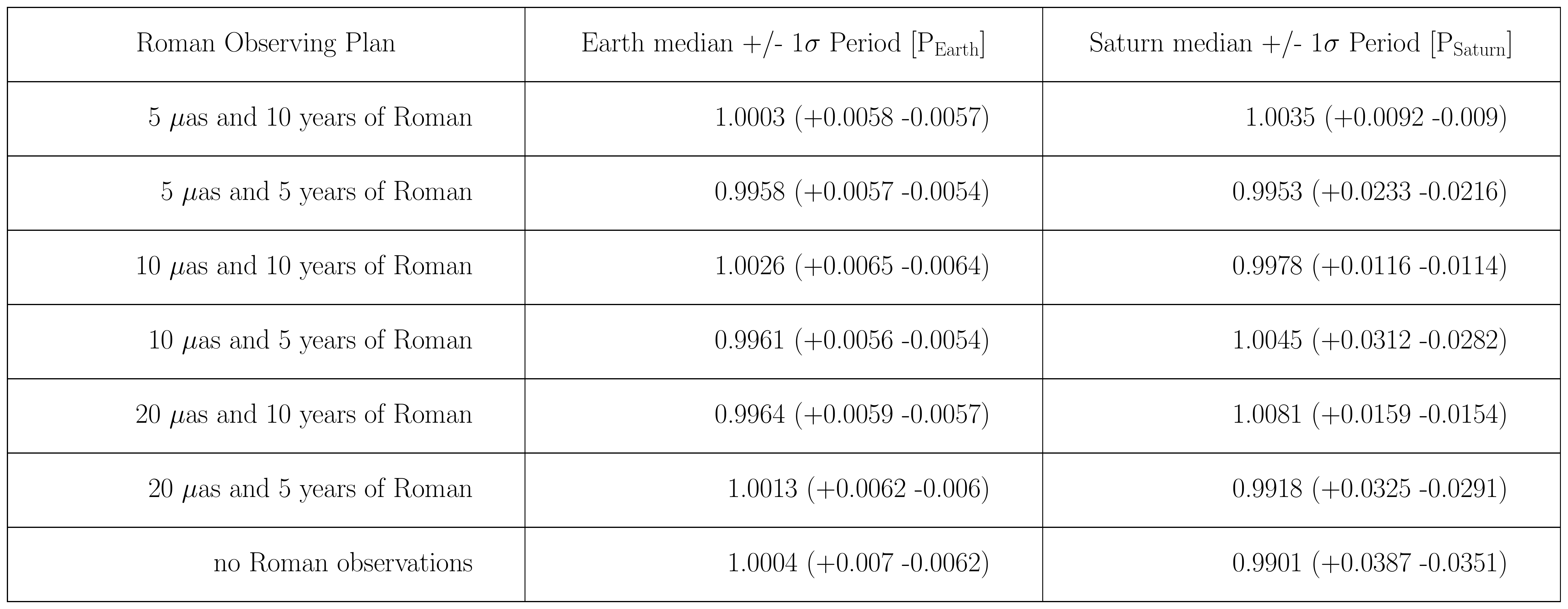}
\caption{Recovered MCMC median and 1$\sigma$ periods for the simulated Earth and Saturn around a Sun at 10 parsecs.}
\label{tab: earth_sat_period}

\end{table}

\begin{table}[!htb]
\centering 
\includegraphics[width=\textwidth]{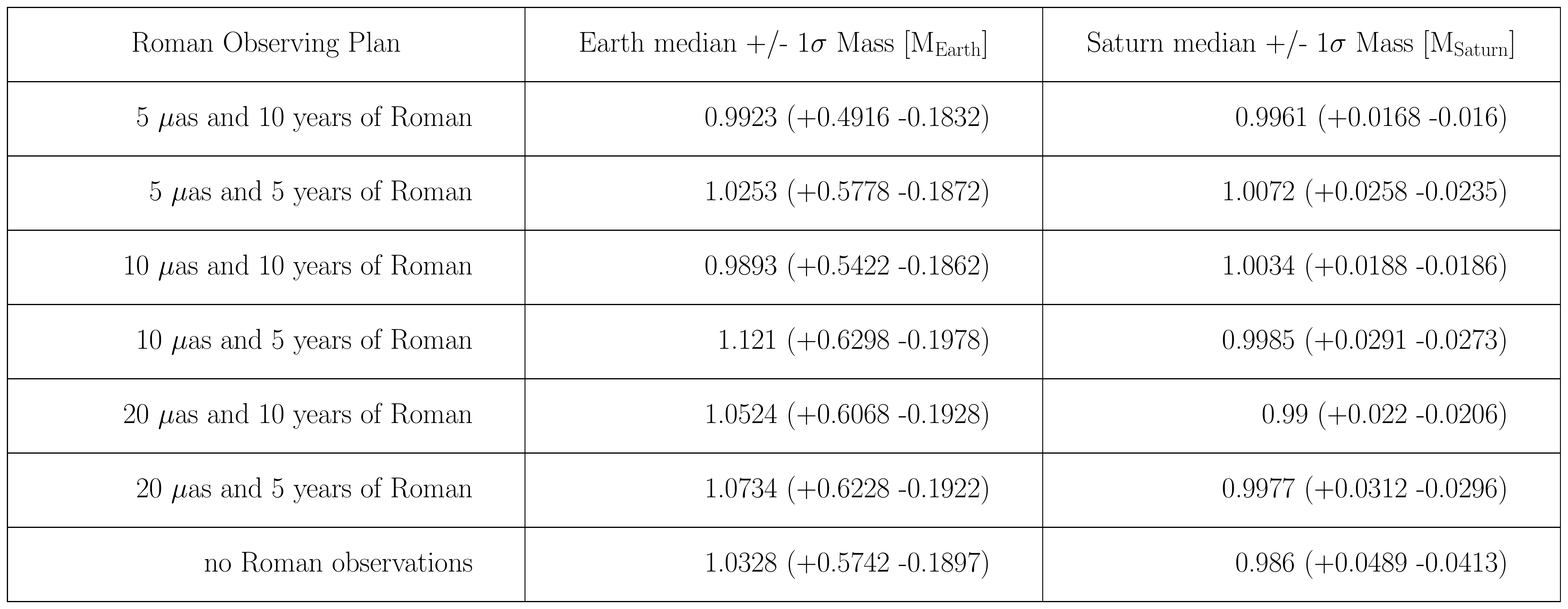}
\caption{recovered MCMC median and 1$\sigma$ masses for the simulated Earth and Saturn around a Sun at 10 parsecs.}
\label{tab: earth_sat_mass}

\end{table}

\begin{figure*}[htb]
\centering 
\includegraphics[width=\textwidth]{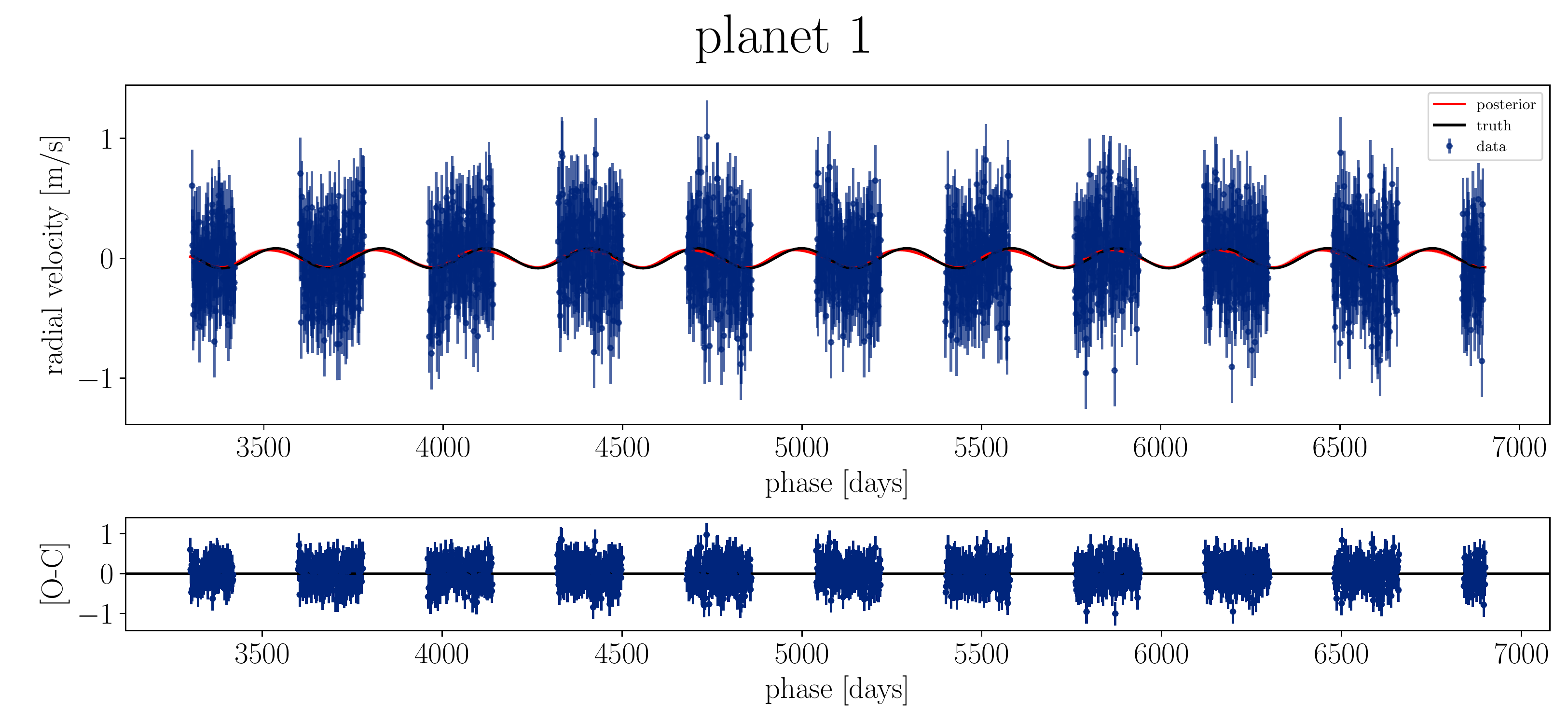}
\caption{Simulated ``true'' Earth-like RVs around Sun-like star. Predicted Terra Hunting Experiment (THE) observations assuming white noise errors of 0.3 m/s, and nightly observations for 180 days per year, for 10 years. And MCMC posterior results for the RVs.}
\label{fig: earth_rvs}

\end{figure*}

\begin{figure*}[htb]
\centering 
\includegraphics[width=\textwidth]{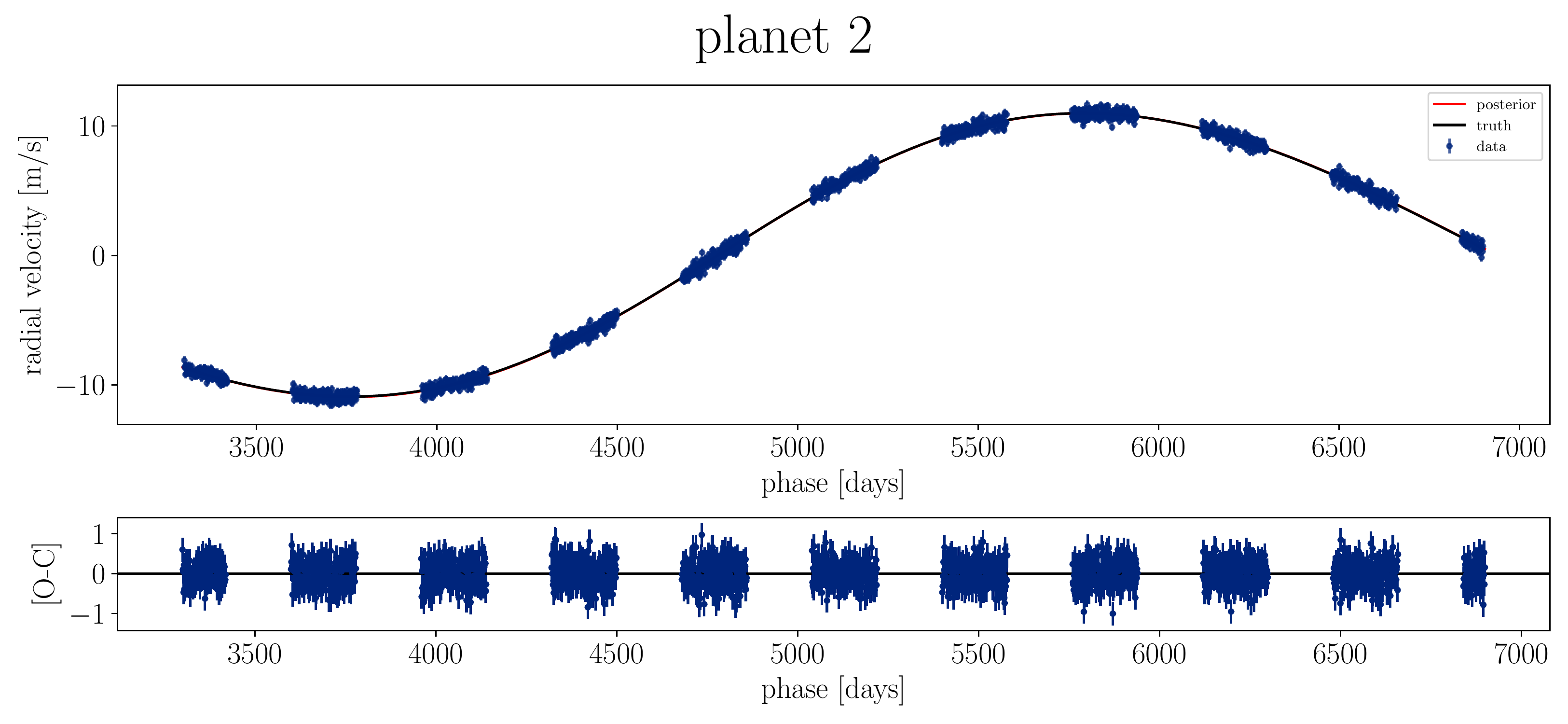}
\caption{Simulated ``true'' RVs for a Jupiter-like exoplanet around Sun-like star. Predicted Terra Hunting Experiment (THE) observations assuming white noise errors of 0.3 m/s, and nightly observations for 180 days per year, for 10 years. And MCMC posterior results for the RVs.}
\label{fig: jupiter_rvs}

\end{figure*}

\begin{figure*}[htb]
\centering 
\includegraphics[width=\textwidth]{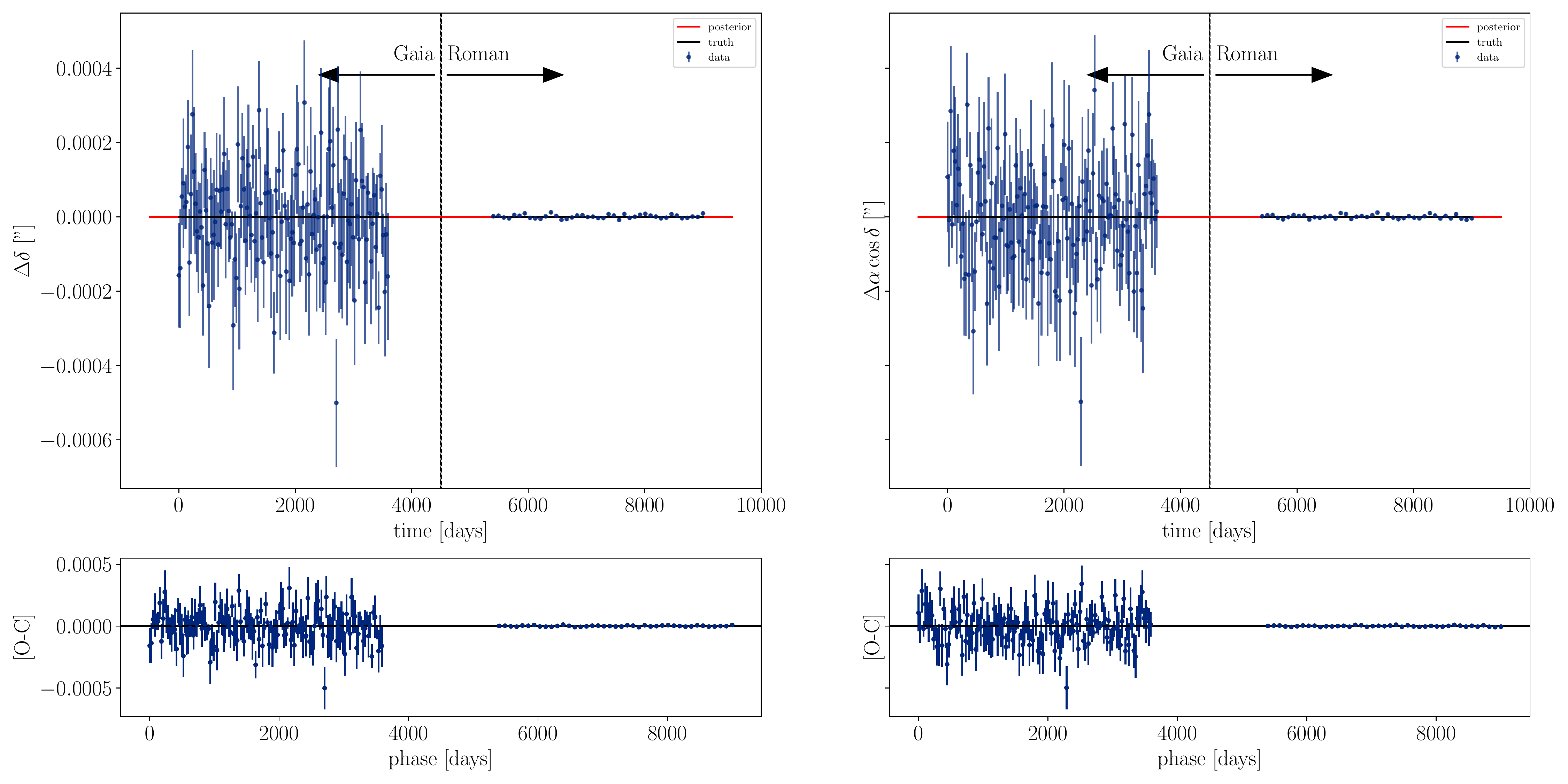}
\caption{Simulated ``true'' RA and Dec astrometry for an Earth-like exoplanet around Sun-like star. Predicted Gaia RA and Dec observations assuming white noise errors of 34.2 $\mu$as along-scan and 171 $\mu$as across-scan with observations every 26 days for 10 years. Predicted Roman RA and Dec observations assuming white noise errors of 5 $\mu$as with observations every 40 days for 10 years. And MCMC posterior results for the RA and Dec.}
\label{fig: earth_astrometry}

\end{figure*}

\begin{figure*}[htb]
\centering 
\includegraphics[width=\textwidth]{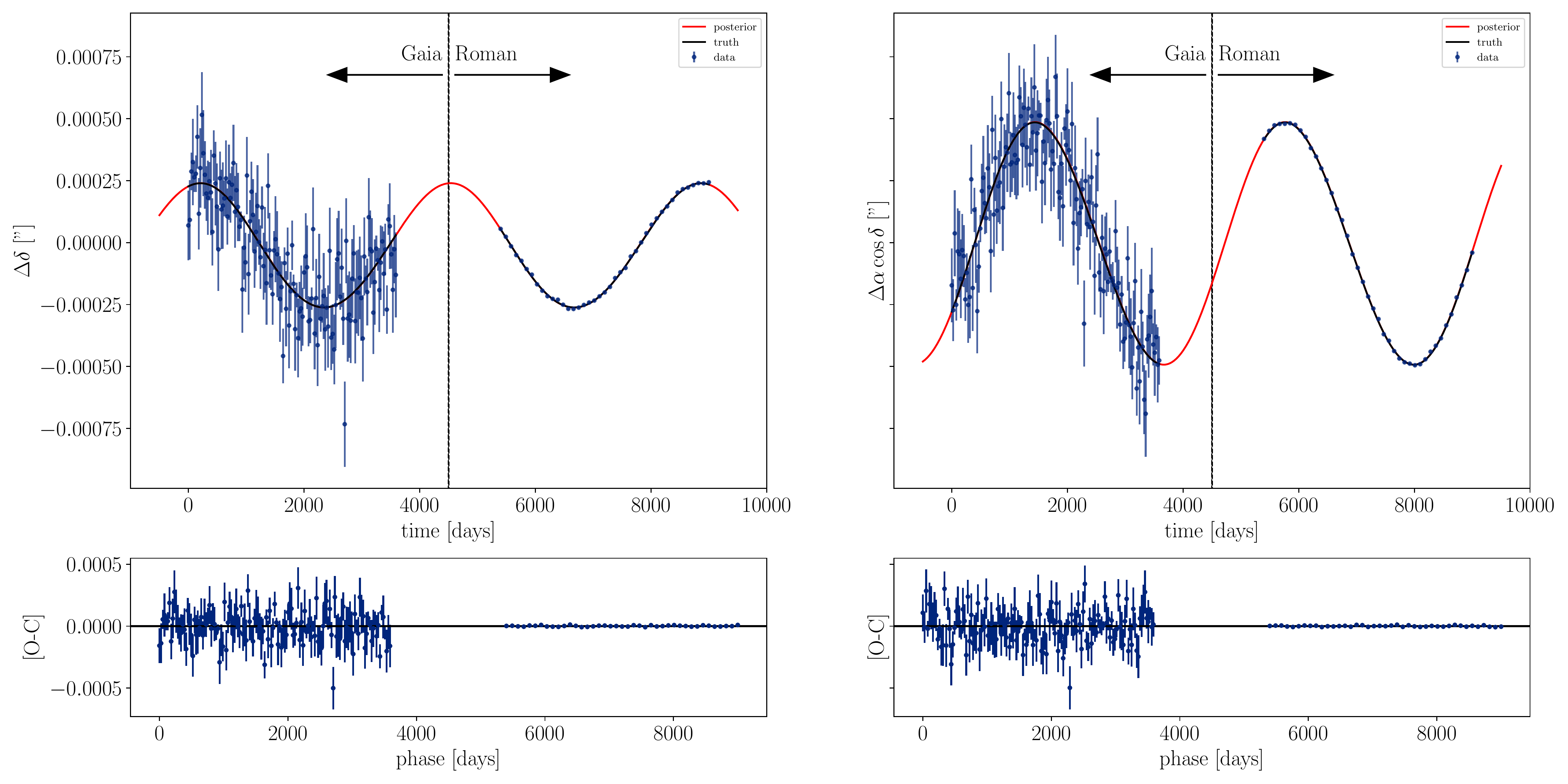}
\caption{Simulated ``true'' RA and Dec astrometry for an Jupiter-like exoplanet around Sun-like star at 10 parsecs. Predicted Gaia RA and Dec observations assuming white noise errors of 60 $\mu$as with observations every 26 days for 10 years. Predicted Roman RA and Dec observations assuming white noise errors of 5 $\mu$as with observations every 40 days for 10 years. And MCMC posterior results for the RA and Dec.}
\label{fig: jupiter_astrometry}

\end{figure*}

\begin{figure*}[htb]
\centering 
\includegraphics[width=\textwidth]{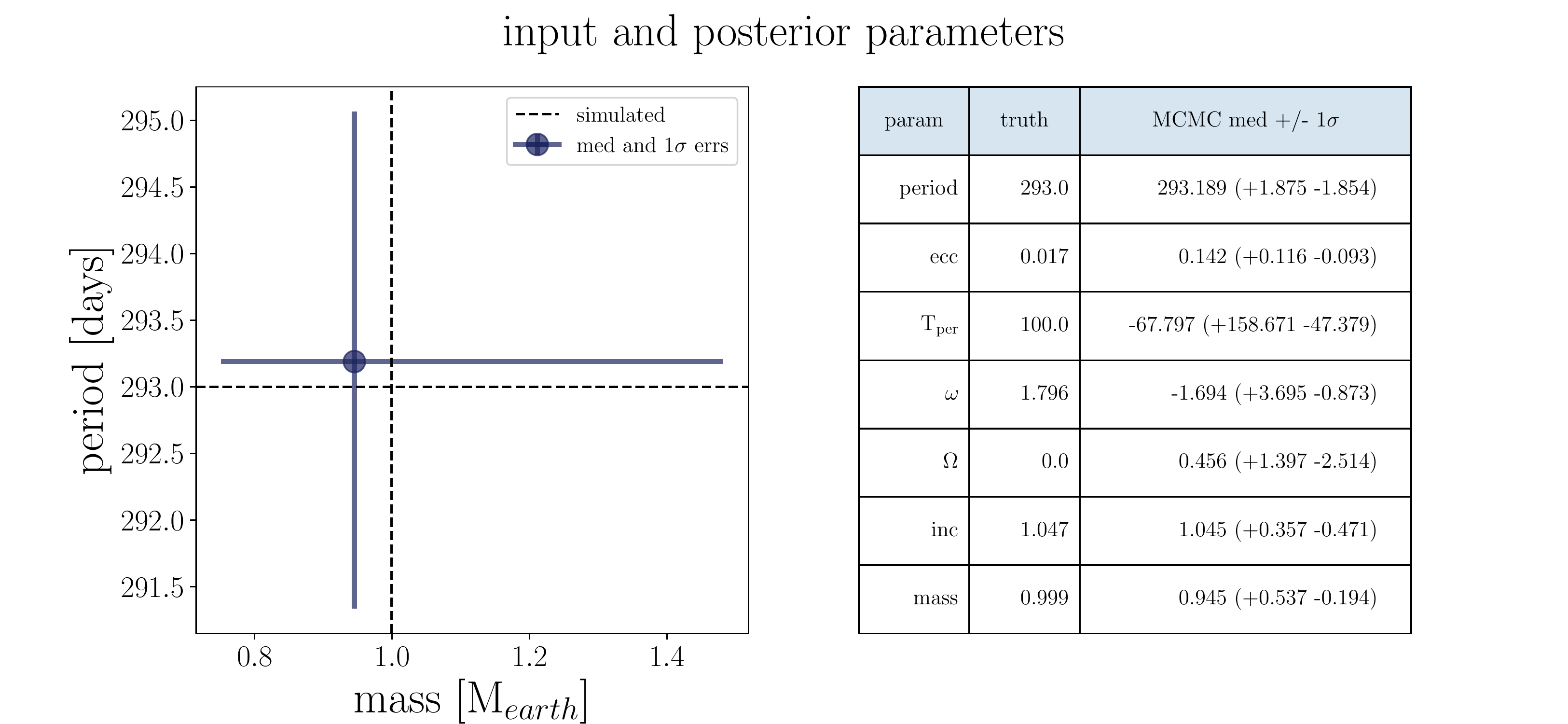}
\caption{$\textbf{Left:}$ Median and 1$\sigma$ MCMC period and mass for a Earth-like exoplanet around a Sun-like star. The black dashed line shows the simulated values. $\textbf{Right:}$ MCMC median and 1$\sigma$ values for exoplanetary and orbital parameters as well as their corresponding simulated ``true'' values. Roman observing program with 10 years of  observations every 40 days and an assumed 1$\sigma$ measurement error of 5 $\mu$as.}
\label{fig: earth_posteriors}

\end{figure*}

\begin{figure*}[htb]
\centering 
\includegraphics[width=\textwidth]{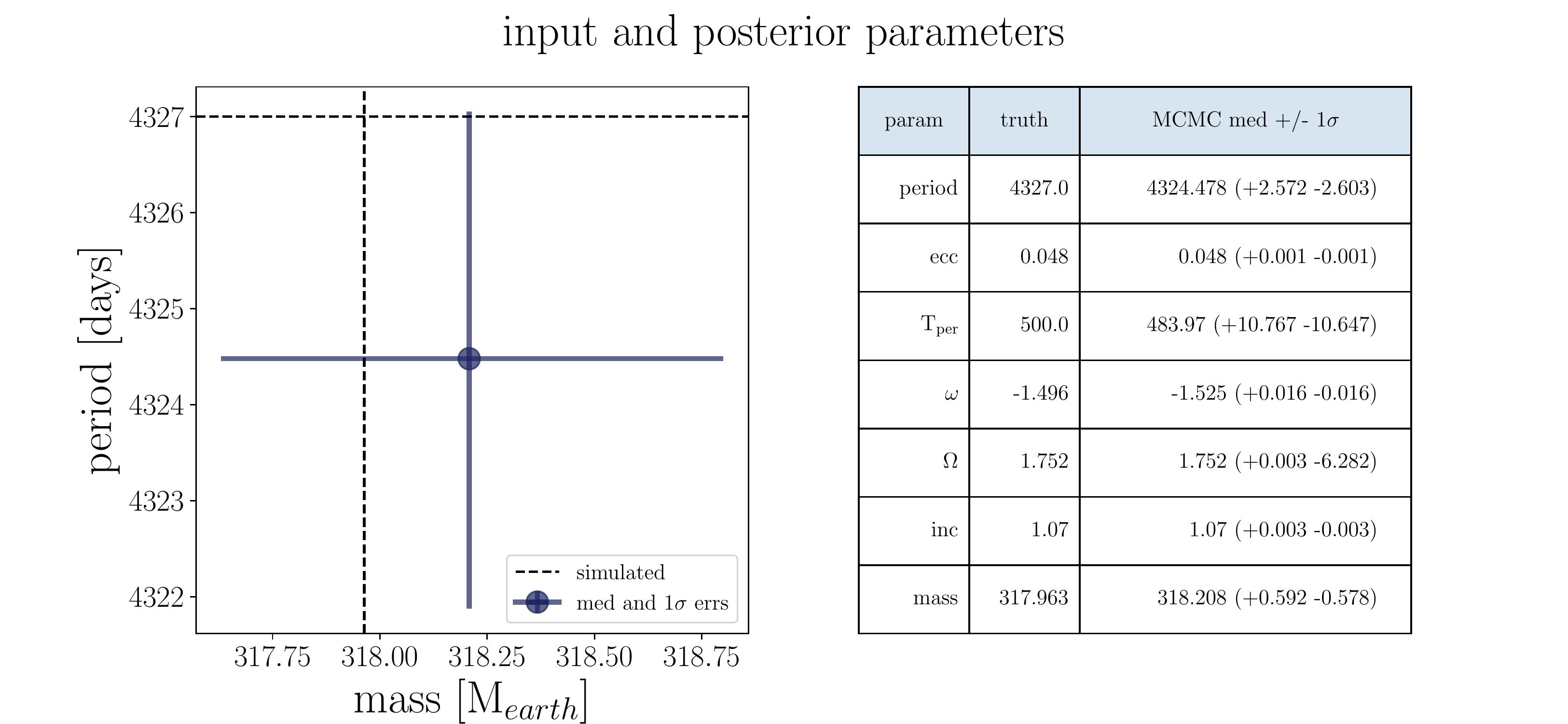}
\caption{$\textbf{Left:}$ Median and 1$\sigma$ MCMC period and mass for a Jupiter-like exoplanet around a Sun-like star. The black dashed line shows the simulated values. $\textbf{Right:}$ MCMC median and 1$\sigma$ values for exoplanetary and orbital parameters as well as their corresponding simulated ``true'' values. Roman observing program with 10 years of  observations every 40 days and an assumed 1$\sigma$ measurement error of 5 $\mu$as.}
\label{fig: jupiter_posteriors}

\end{figure*}

\newpage
In what follows, we present one example of a simulated and modeled Earth and Jupiter exoplanetary system. For this run, we simulated 10 years of a Roman observing with white noise of 5 $\mu$as per astrometric observation. Figure \ref{fig: earth_rvs} shows the simulated RV data, the true orbit, and the MCMC posterior orbit for Earth. Figure \ref{fig: jupiter_rvs} shows the simulated RV data, the true orbit, and the MCMC posterior orbit for Jupiter. Figure \ref{fig: earth_astrometry} shows the simulated RV data, the true orbit, and the MCMC posterior orbit for Earth. Figure \ref{fig: jupiter_astrometry} shows the simulated RV data, the true orbit, and the MCMC posterior orbit for Jupiter. Figure \ref{fig: earth_posteriors} shows the MCMC median and 1$\sigma$ values for some key planetary parameters for Earth, as well as the ``true'' simulated values for Earth. Figure \ref{fig: jupiter_posteriors} shows the MCMC median and 1$\sigma$ values for some key planetary parameters for Earth, as well as the ``true'' simulated values for Jupiter. Finally, Figure \ref{fig: corner} shows a corner plot for all model parameters, in which we can see several covariances and dependencies of the different model parameters as discussed in Section \ref{discussion}.

\newpage

\end{document}